\documentclass[fleqn, 10pt]{wlscirep}
\usepackage[utf8]{inputenc}
\usepackage[T1]{fontenc}
\usepackage{algorithm}
 \usepackage{graphicx}
 \usepackage{amsthm}
 \usepackage{subcaption}
 \usepackage{comment}
\usepackage{algpseudocode} 
\usepackage{siunitx}
\usepackage{amsmath}
\usepackage{dcolumn}
\usepackage{multirow}%
\usepackage{xcolor}%
\usepackage{textcomp}%
\usepackage{manyfoot}%
\usepackage{braket}
\usepackage[section]{placeins}
\usepackage{bm}
\usepackage{ulem, censor}

\usepackage{caption}
\usepackage{lipsum}
\usepackage{tikz}
\usepackage{amsfonts}
\usepackage{amssymb}

\DeclareMathOperator*{\argmax}{arg\,max}
\DeclareMathOperator*{\argmin}{arg\,min}

 \newtheorem{theorem}{Theorem}

\usepackage{soul}
\usepackage{amsfonts}
\title{Greedy Gradient-free Adaptive Variational Quantum Algorithms on a Noisy Intermediate Scale Quantum Computer}

\author[1,2]{César Feniou}
\author[3]{Muhammad Hassan}
\author[2]{Baptiste Claudon}
\author[2]{Axel Courtat}
\author[1]{Olivier Adjoua}
\author[3]{Yvon Maday}
\author[1,2,*]{Jean-Philip Piquemal}

\affil[1]{Sorbonne Universit\'e, Laboratoire de Chimie Théorique (UMR-7616-CNRS), F-75005 Paris, France}
\affil[2]{Qubit Pharmaceuticals, Advanced Research Department, Paris, France}
\affil[3]{Sorbonne Université, Université Paris Cité, CNRS, INRIA, Laboratoire Jacques-Louis Lions (LJLL), F-75005 Paris, France}

\affil[*]{jean-philip.piquemal@sorbonne-universite.fr}

\begin{abstract}
Hybrid quantum-classical adaptive Variational Quantum Eigensolvers (VQE) hold the potential to outperform classical computing for simulating many-body quantum systems. However, practical implementations on current quantum processing units (QPUs) are challenging due to the noisy evaluation of a polynomially scaling number of observables, undertaken for operator selection and high-dimensional cost function optimization. We introduce an adaptive algorithm using analytic, gradient-free optimization, called Greedy Gradient-free Adaptive VQE (GGA-VQE). In addition to demonstrating the algorithm’s improved resilience to statistical sampling noise in the computation of simple molecular ground states, we execute GGA-VQE on a 25-qubit error-mitigated QPU by computing the ground state of a 25-body Ising model. Although hardware noise on the QPU produces inaccurate energies, our implementation outputs a parameterized quantum circuit yielding a favorable ground-state approximation. We demonstrate this by retrieving the parameterized operators calculated on the QPU and evaluating the resulting ansatz wave-function via noiseless emulation (i.e., hybrid observable measurement).

\end{abstract}

\begin{document}

\flushbottom
\maketitle
%
%
\thispagestyle{empty}
\vspace{-10mm}
\section{Introduction}\label{sec1}

Quantum computing has gained considerable interest due to its potential to solve complex computational problems that are intractable on classical devices. Finding the ground state of a many-body quantum system is one such problem as it suffers from an exponentially scaling complexity in the system size \cite{feynman2018simulating}. Quantum computing provides an appealing solution as it allows, in principle, the encoding of the exponentially scaling many-body wave-function onto a linearly scaling qubit register. In the context of quantum chemistry, extensive efforts have been made to develop quantum algorithms for ground and excited state preparation of molecular systems with the goal of ultimately surpassing classical techniques \cite{cao2019quantum, mcardle2020quantum}. In order to be able to take advantage of near-term quantum devices in the noisy intermediate-scale quantum (NISQ) era, emphasis has been placed on the development of hybrid quantum-classical algorithms such as the Variational Quantum Eigensolver (VQE) which incorporates a quantum subroutine within a classical optimization loop thereby allowing for the most costly aspect of the optimisation routine (namely, measuring the expectation value of the Hamiltonian) to be be performed on the quantum device \cite{peruzzo2014variational}.

The core idea of the variational quantum eigensolver is to generate a parameterised wave-function, known as the ansatz, and then variationally tune this ansatz so as to minimise the expectation value of some relevant Hermitian operator, typically the system Hamiltonian. The fundamental challenge in implementing the VQE methodology on NISQ devices is thus to construct an ansatz wave-function that can accurately describe the ground-state of the Hermitian operator under study and, at the same time, can be represented on shallow quantum circuits which are not dominated by noise. Most commonly used VQEs for quantum chemistry are so-called ``fixed-ansatz'' methods wherein the ansatz wave-function consists of a predetermined product of parametrised unitary operators acting on an initial (usually Hartree-Fock reference) state \cite{tilly2022variational, romero2018strategies, ryabinkin2018qubit, kandala2017hardware, lee2018generalized, mizukami2020orbital}. Whether hardware-efficient or chemically-inspired, these "fixed" ansatz methods have a limited accuracy and do not provide a route for exact simulations of strongly correlated systems on near-term quantum hardware \cite{romero2018strategies, kandala2017hardware}. Since fixed ansätze are by definition system-agnostic, they are also likely to contain superfluous operators that do not contribute to a better approximation of the ground state of the Hermitian operator under study \cite{grimsley2019trotterized}. Such redundant operators needlessly increase the length of the ansatz circuit as well as the number of variational parameters to be tuned, both of which are serious problems for NISQ-era devices.

Unsurprisingly therefore, recent works have proposed iterative VQE protocols that construct system-tailored ans\"atze using some kind of quasi-greedy strategy. The ADAPT-VQE algorithm \cite{grimsley2019adaptive} has made a notable impact in the field by demonstrating a significant reduction in the redundant terms in the ansatz circuits for a range of molecules, thus enhancing the accuracy and efficiency of the VQE. At it's core, the ADAPT-VQE algorithm consists of two steps:
\begin{description}
    \item[Step 1] At iteration $m$, we are given a parameterised ansatz wave-function $\ket{\Psi^{(m-1)}}:=\ket{\Psi^{(m-1)}(\theta_{m-1}^{(m-1)}, \ldots, \theta_{1}^{(m-1)})}$ (see Equation \eqref{eq:Yvon_new} in Section~\ref{sec:Adapt} for the precise form of $\ket{\Psi^{(m-1)}}$). A new ansatz wave-function is constructed by carefully selecting a parameterised unitary operator from a pre-selected operator pool and appending this parameterised unitary operator to $\ket{\Psi^{(m-1)}}$. More precisely, given a pool $\mathbb{U}$ of parameterised unitary operators and a Hermitian operator $\widehat{A}$ whose ground state is being prepared, the ADAPT-VQE selection criterion consists of identifying the unitary operator $\mathcal{U}^* \in \mathbb{U}$ such that
 \begin{align}\label{eq:adapt_vqe_criterion}
     \mathcal{U}^*= \underset{\mathcal{U} \in \mathbb{U}}{\text{argmax}} \left \vert \frac{d}{d\theta} \Big \langle \Psi^{(m-1)}(\theta_{m-1}^{(m-1)}, \ldots, \theta_{1}^{(m-1)})\left \vert \mathcal{U}(\theta)^\dagger \widehat{A}  \mathcal{U}(\theta)\right \vert \Psi^{(m-1)}(\theta_{m-1}^{(m-1)}, \ldots, \theta_{1}^{(m-1)}) \Big \rangle  \Big \vert_{\theta=0} \right \vert.
 \end{align}
This results in an auxiliary wave-function of the form
\[\ket{{\Psi}^{(m)}(\theta_{m}, \theta_{m-1}^{(m-1)}, \ldots, \theta_{1}^{(m-1)})}:= \mathcal{U}^*(\theta_{m})\ket{\Psi^{(m-1)}(\theta_{m-1}^{(m-1)}, \ldots, \theta_{1}^{(m-1)})}.\] 


Note that, at this stage, $\theta_{m}$ is a free parameter whose value has not yet been fixed.

    \item[Step 2] We now perform a \emph{global optimisation} over all parameters $\theta_1^{(m-1)}, \ldots, \theta_{m-1}^{(m-1)}, \theta_{m}$ of the expectation value of the Hermitian operator under study. More precisely, we solve the $m$-dimensional optimisation problem
    \begin{align*}
    (\theta_1^{(m)}, \ldots, \theta_{m-1}^{(m)}, \theta_m^{(m)}):=\underset{\theta_1, \ldots, \theta_{m-1}, \theta_{m}}{\operatorname{argmin}}  \Big \langle {\Psi}^{(m)}(\theta_{m}, \theta_{m-1}, \ldots, \theta_{1})\left \vert\; \widehat{A} \;\right \vert {\Psi}^{(m)}(\theta_{m}, \theta_{m-1}, \ldots, \theta_{1})\Big \rangle,
    \end{align*}
    and define the new ansatz wave-function as $\ket{\Psi^{(m)}}:=\ket{\Psi^{(m)}(\theta_{m}^{(m)}, \theta_{m-1}^{(m)}, \ldots, \theta_{1}^{(m)})}.$ Note that for notational convenience, when explaining such adaptive algorithms in the sequel, we will drop the parameter superscript indices and write $\ket{\Psi^{(m)}}=\ket{\Psi^{(m)}(\theta_{m}', \theta_{m-1}', \ldots, \theta_{1}' )}$ with the understanding that each parameter $\theta_j'$ may change from iteration to iteration. 
\end{description}

\begin{figure}[H]
    \centering
    \begin{subfigure}[b]{0.49\textwidth}
        \centering
        \includegraphics[width=\textwidth]{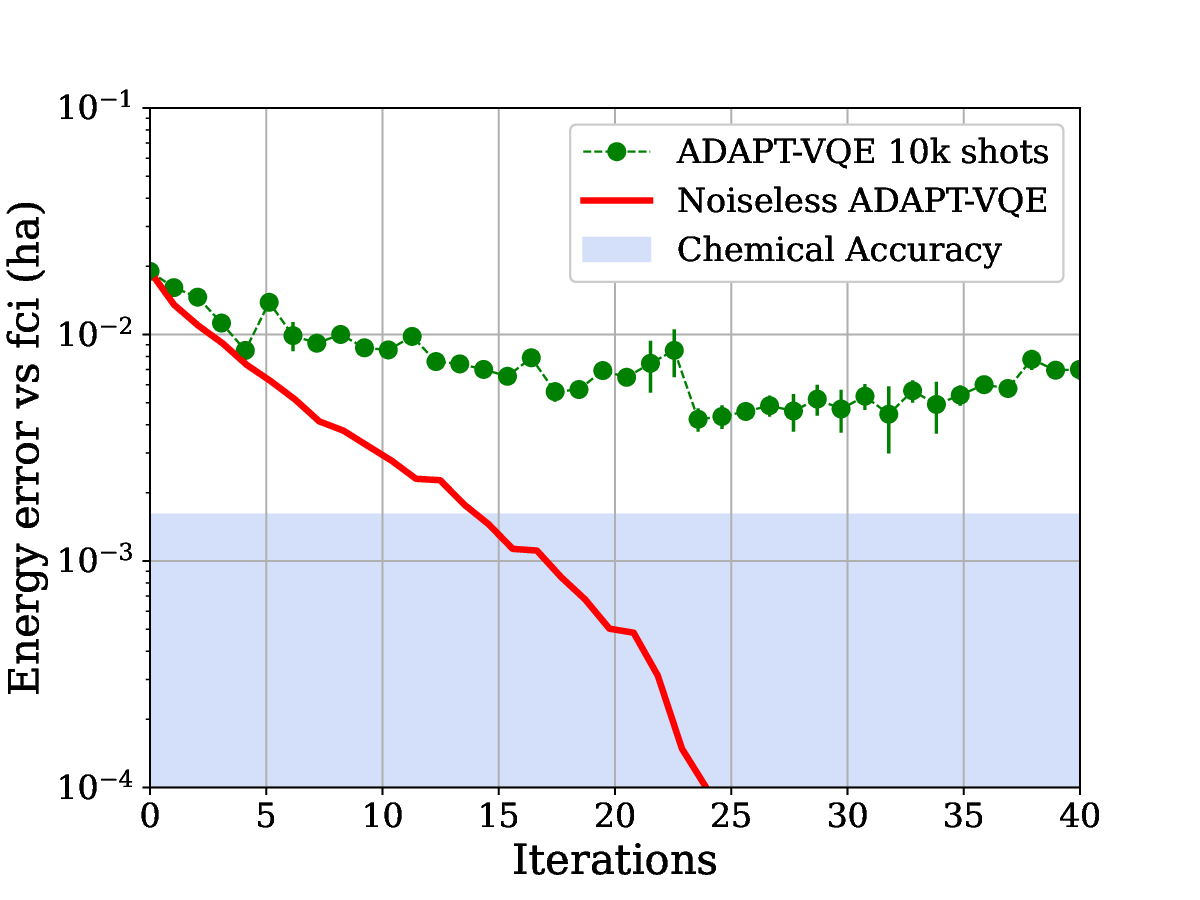}
        \caption{H\(_2\)O}
        \label{fig:h2oadapt}
    \end{subfigure}
    \begin{subfigure}[b]{0.49\textwidth}
        \centering
        \includegraphics[width=\textwidth]{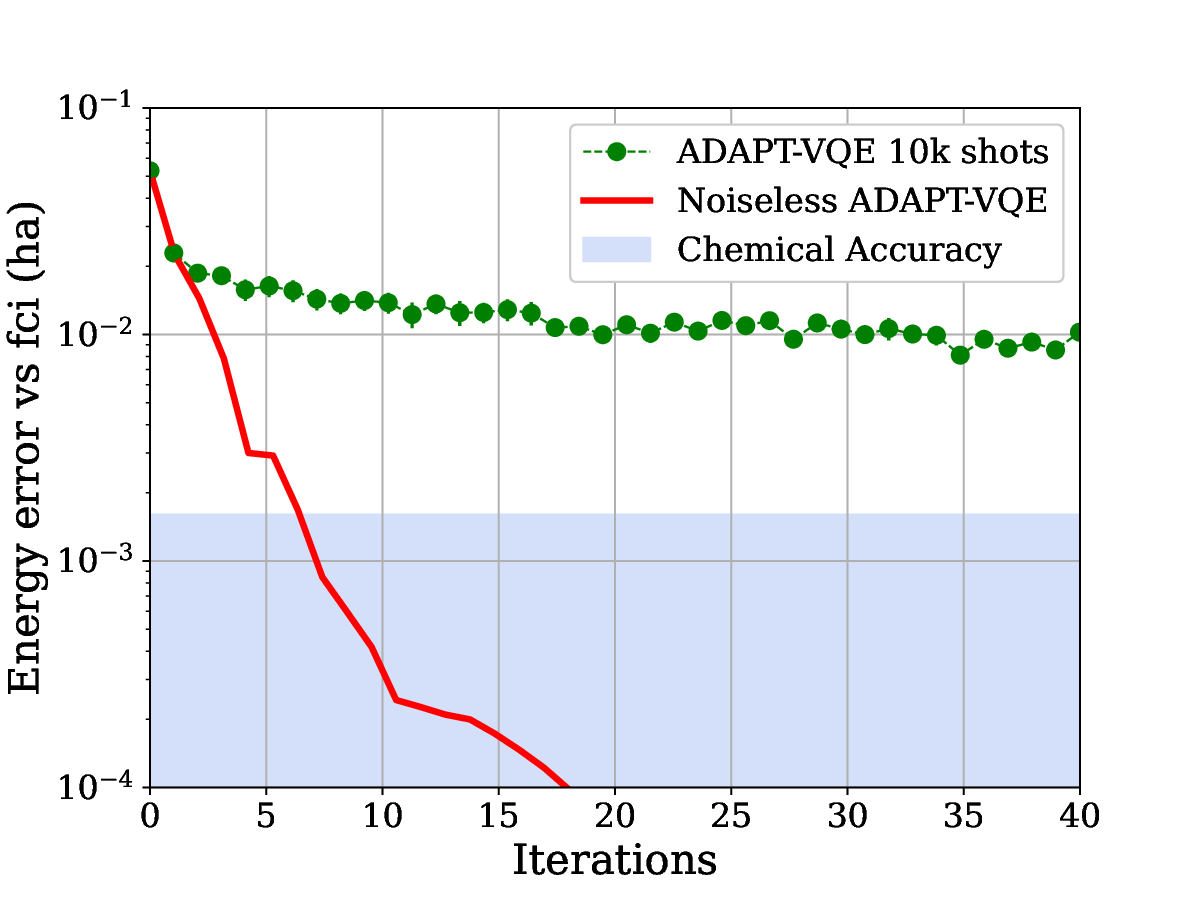}
        \caption{LiH}
        \label{fig:lihadapt}
    \end{subfigure}
    \caption{The effect of introducing shot noise on the energy convergence of ADAPT-VQE simulations for H\(_2\)O and LiH molecules. The numerical parameters used to obtain these results are the same as those used for the numerical experiments in Section \ref{sec:numerics_1} and can be found at the beginning of Section \ref{sec:results}.}
    \label{fig:adapt}
\end{figure}

Each of the above two steps presents challenges if one wishes to implement ADAPT-VQE on a practical NISQ device. Indeed, the operator selection procedure (\textbf{Step 1}) involves computing gradients of the expectation value of the Hamiltonian for every choice of operator in the operator pool $\mathbb{U}$, which typically requires tens of thousands of extremely noisy measurements on the quantum device. The global optimisation procedure (\textbf{Step 2}) wherein the ansatz wave-function is variationally tuned presents a similar problem because the underlying cost function-- since it arises from evaluations on a NISQ device-- is non-linear, high-dimensional and noisy, thus often rendering the associated optimisation problem computationally intractable \cite{astrakhantsev2022phenomenological}. These issues are illustrated in Figure \ref{fig:adapt} where we compare the results of noiseless and noisy (in the form of statistical noise using 10,000 shots on an HPC emulator) ADAPT-VQE simulations for the dynamically correlated H$_2$O and LiH molecules. For both molecules, we see that the algorithm correctly recovers the exact ground state energy of the molecule to high accuracy in the \emph{noiseless} regime. On the other hand, introducing measurement noise immediately reduces the quality of the ADAPT-VQE output, and the algorithm stagnates well-above the chemical accuracy threshold of $1$ milliHartree despite a moderate number of iterations.

It is therefore not surprising that, despite some partial attempts \cite{rossmannek2023quantum}, full implementations of ADAPT-VQE-type algorithms on the current generation of quantum hardware are yet to be achieved. Indeed, a recent study \cite{dalton2024quantifying} suggests that quantum gate errors need to be reduced by orders of magnitude before current VQEs can be expected to bring a quantum advantage. Interestingly, while a great deal of effort has been devoted to developing various improvements of the original ADAPT procedure, a majority of this research has focused on further compactifying the adaptively generated ansatz wave-functions, thus reducing the quantum circuit depth and the number of CNOT gates required to represent the ansätze (see \cite{yordanov2021qubit, tang2021qubit, anastasiou2022tetris, burton2023exact, feniou2023overlapadaptvqe}). Unfortunately, these improvements either do not deal with, or even worse, come at the expense of further increasing the number of measurements on the quantum device required to perform the iterative procedure. Comparatively fewer articles in the literature have focused on decreasing the quantum measurement overhead required to implement ADAPT-VQE-type methods on current NISQ devices but these include the notable advances \cite{choi2022improving, loaiza2022reducing, yen2023deterministic, nykanen2022mitigating} which outline several clever ways to reduce the number of measurements required to evaluate the expectation value of the Hamiltonian. Concerning specifically \textbf{Step 1} in the ADAPT-VQE algorithm, the contribution \cite{anastasiou2023really} introduces a novel strategy for simultaneously evaluating the gradients of the expectation value of the Hamiltonian for multiple operators in the operator pool, thereby resulting in a drastic decrease in the number of quantum measurements required for operator selection. Similarly, the article \cite{liu2021efficient} proposes the use of adaptive ansatz wave-functions using reduced density matrices, which also results in a considerable reduction in the quantum measurement overhead for operator selection. Despite the aforementioned contributions, the gap between the quantum resources afforded by the current generation of quantum hardware and those required by these improved adaptive algorithms is yet to be bridged.

The goal of the present contribution is to study the feasibility of NISQ implementations of adaptive quantum algorithms by implementing a simplification of \textbf{Step 2} above, namely, the global optimisation of a non-linear, high-dimensional noisy cost function. We do so by introducing a noise-resistant and resource-efficient, greedy gradient-free adaptive variational quantum algorithm that is motivated by the gradient-free, analytical optimisation approaches that have been proposed in the VQE literature on `fixed-ansatz' methods. Indeed, it is well-known that the expectation value of a Hermitian operator with respect to an ansatz wave-function $\Psi(\theta)$ parametrised by a single quantum gate is simply an elementary trigonometric function of $\theta$ \cite{nakanishi2020sequential, ostaszewski2021structure, Wada_2022, watanabe2023optimizing, crooks2019gradients, izmaylov2021analytic, wierichs2022general}. Thus, for a wave-function parametrised by a single rotation gate, only two measurements of the expectation value for judiciously chosen values of $\theta$, allow an exact reconstruction of the full expectation value as a parametrised function of $\theta$. Using this insight, we propose to replace the conventional ADAPT-VQE operator selection criterion with a newly developed gradient-free energy sorting approach that allows us to identify both the \emph{the locally optimal} parametrised unitary operator and the associated optimal angle, which when appended to the current ansatz wave-function, will produce a new ansatz wave-function with the biggest drop in expectation value. In other words, in contrast to the gradient-based operator selection \textbf{Step \ref{eq:adapt_vqe_criterion}} followed by global optimisation procedure used in ADAPT-VQE, our adaptive algorithm selects, in one shot, both a locally optimal unitary operator $\mathcal{U}^*$ and associated optimal angle $\theta_{m+1}'$ that satisfy
\begin{align}\label{eq:new_criterion}
(\mathcal{U}^*, \theta_{m+1}') = \underset{\mathcal{U} \in \mathbb{U}, \theta \in [-\pi, \pi)}{\text{argmin}}  
    \langle \Psi^{(m)}(\theta_{m}', \ldots, \theta_{1}')\left \vert \mathcal{U}(\theta)^\dagger \widehat{A}  \mathcal{U}(\theta)\right \vert \Psi^{(m)}(\theta_{m}', \ldots, \theta_{1}')\rangle.
\end{align}

As we discuss in Section \ref{sec:Energy_Sorted} below, the one-dimensional objective functions appearing in the optimisation problem \eqref{eq:new_criterion}, also known as \emph{landscape} functions, can be expressed as analytical functions of $\theta$ for any unitary operator $\mathcal{U}$ belonging to popular choices of operator pools. These analytical landscape functions can be determined explicitly using a fixed number of measurements that depends on the nature of the operator pool and the Hermitian operator $\widehat{A}$ which can allow us to determine simultaneously both the best unitary operator $\mathcal{U}^*$ and the optimal angle $\theta_{m+1}'$ that should be used to update the current ansatz wave-function. By iteratively growing an ansatz wave-function using only such locally optimal parametrised unitary operators and not re-optimising the ``frozen-core'' of the previous ansatz, we are able to avoid the \emph{global optimisation} of a multi-dimensional noisy objective function. We refer to this adaptive algorithm as the greedy, gradient-free adaptive variational quantum eigensolver or GGA-VQE algorithm for short (see Section~\ref{sec:Energy_Sorted} below for a detailed description). While gradient-free, analytical optimisation approaches for VQEs have certainly been explored in the `fixed-ansatz' literature \cite{nakanishi2020sequential, ostaszewski2021structure, Wada_2022, watanabe2023optimizing} and energy-sorting algorithms to improve the ADAPT-VQE operator selection criterion have also been proposed \cite{yordanov2022molecular, fan2021circuit}, to the best of our knowledge, our work is the first attempt to combine both approaches and develop a greedy gradient-free adaptive variational quantum VQE.

It is important to point out that we do not expect our proposed algorithm to be able to produce highly accurate approximations of the ground state energies of strongly correlated molecules without further post-processing (considerations in this direction are discussed in Section \ref{sec:d-dimen} in the appendix). However, given that adaptive VQEs struggle to produce chemically accurate approximations of the ground states of even weakly correlated molecules on NISQ devices, we feel that the goal of tackling strongly correlated systems on the current generation of NISQ devices is too ambitious. Instead our aim is to simply develop a noise-resistant adaptive VQE that has the potential to handle accessible molecular systems on near-term quantum devices.

To explore the suitability of our proposal, we compare our GGA-VQE algorithm with ADAPT-VQE for some simple molecular systems using an High-Performance Computing (HPC) emulator that includes statistical noise from sampling. For such weakly correlated system, we see a marked improvement over the ADAPT-VQE results. Encouraged by the results of these HPC simulations, we next consider the ground state preparation of an open boundary, one-dimensional transverse-field Ising model on a real quantum device. We show that for Ising Hamiltonians of this nature, using a minimal hardware-efficient operator pool \cite{tang2021qubit}, each iteration of the GGA-VQE algorithm requires measuring only \emph{five} observables on quantum circuits, regardless of the system size (i.e., the number of qubits involved). As a proof of concept, we run the GGA-VQE algorithm for such an Ising model on a 25-qubit register on a state-of-the-art, trapped ion quantum computer and successfully achieve a ground state fidelity of over $98\%$. 
\section{Results}\label{sec:results}

The algorithmic procedures in this research have been performed using an in-house code developed within (i) the IBM Qiskit SDK\cite{javadi2024quantum} for the classical simulations with statistical noise, (ii) our internal Hyperion-1 multi-GPU-accelerated quantum emulator\cite{adjoua2023} using a single NVIDIA DGX A100 node, and (iii) the Amazon Braket SDK\cite{gonzalez2021cloud} for quantum hardware implementations. The quantum computations have been executed on an IonQ Aria 25-qubit trapped-ion quantum computer which incorporates built-in error mitigation techniques, and each observable is evaluated using 2500 shots. Every ADAPT-VQE computation was performed using the gradient-free COBYLA optimizer -- with default Scipy options: final optimization accuracy tolerance (tol) of $1.10^{-4}$, constraint violation tolerance (catol) of $2.10^{-4}$, and a maximum of 1000 inner iterations -- since it yielded the most favourable results; the Broyden–Fletcher–Goldfarb–Shanno (BFGS) algorithm which is the method-of-choice in the literature led to very poor results in our simulations and was therefore avoided (see Figure \ref{fig:bfgs}). All molecular systems are simulated with a STO-3G minimal basis set. For H$_2$O and HF, the 1s orbital is kept frozen which results in active spaces of 12 and 10 spin-orbitals, respectively. For H$_6$ and LiH, all orbitals are active, resulting in 12 spin-orbitals each.

\subsection{The GGA-VQE algorithm applied to Molecular Systems with Measurement Noise}\label{sec:numerics_1}

\begin{figure}[H]
  \centering
  \begin{subfigure}[b]{0.48\textwidth}
    \centering
    \includegraphics[width=\textwidth]{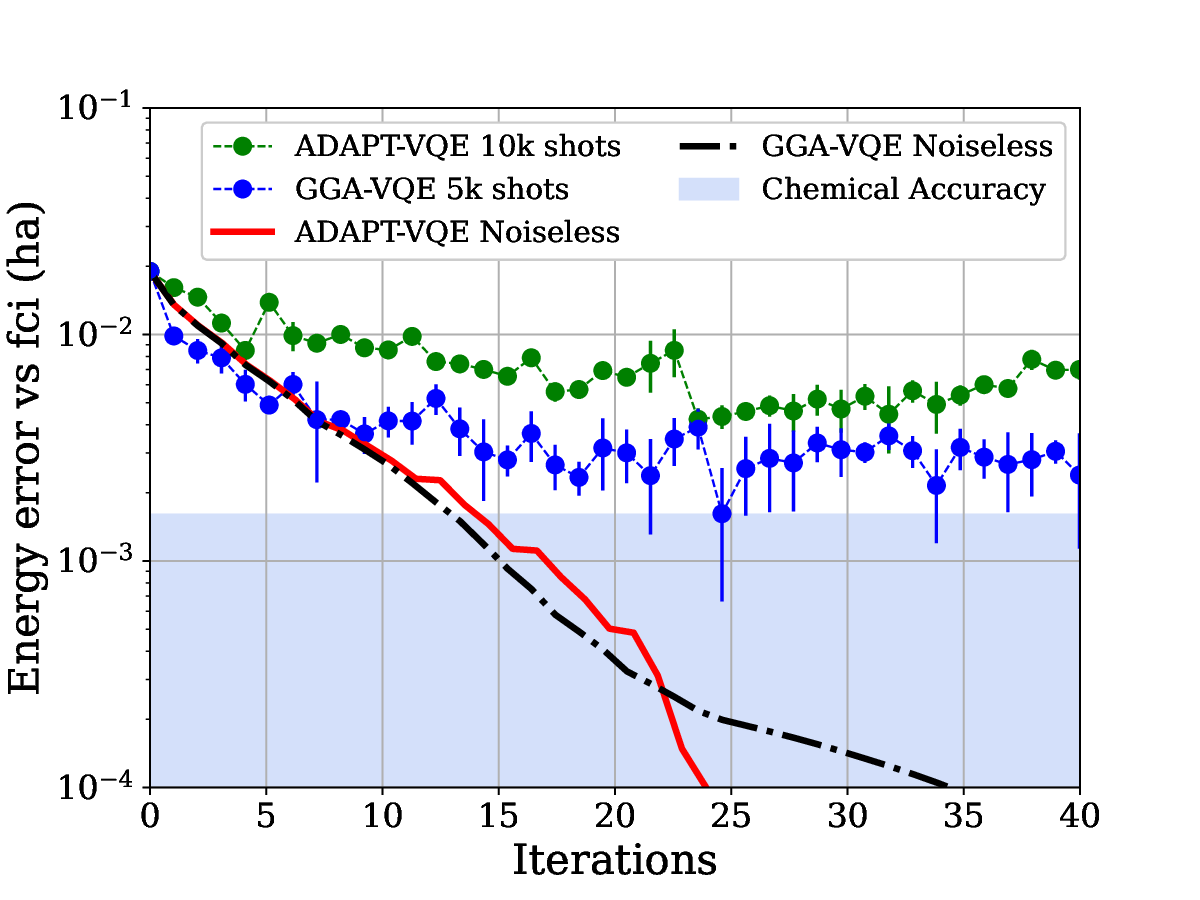}
    \subcaption{H$_2$O molecule.}
  \end{subfigure}
    \begin{subfigure}[b]{0.48\textwidth}
    \centering
    \includegraphics[width=\textwidth]{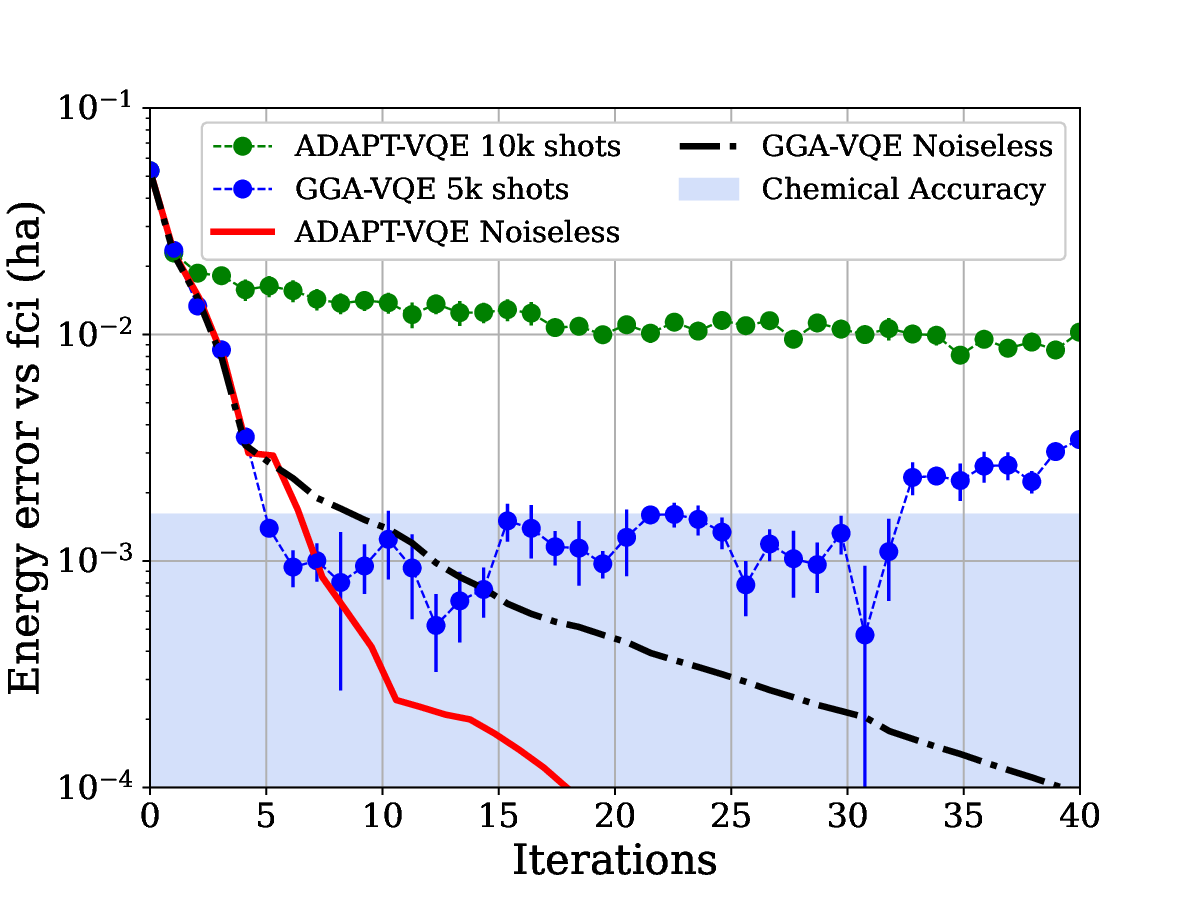}
    \subcaption{LiH molecule.}
  \end{subfigure}
  \caption{Comparison of the GGA-VQE and ADAPT-VQE algorithms for the ground state energy of an H$_2$O molecule (left) and an LiH molecule (right) at equilibrium geometries. Both plots represent the energy convergence as a function of the number of iterations of the algorithms. The shaded blue region indicates chemical accuracy at $10^{-3}$ Hartree. Both simulations involved 12 qubits.}
  \label{fig:GGA_1}
\end{figure}

For our first set of numerical experiments, we apply the GGA-VQE algorithm described in Section \ref{sec:Energy_Sorted} to some weakly correlated molecular systems, and with statistical noise due to the sampling of expectation values from a limited number of shots. Specifically, the expectation value of \emph{each} Pauli string (for instance those arising from the Jordan-Wigner transformation of the Hamiltonian) was measured using 10,000 shots for the ADAPT-VQE and 5,000 shots for the GGA-VQE algorithms respectively. Using half as many shots for observable evaluations in the ADAPT-VQE routine is motivated by the fact that ADAPT-VQE requires only two energy evaluations of the chemical Hamiltonian per operator whereas GGA-VQE requires four energy evaluations of the chemical Hamiltonian per operator in the QEB pool. Since ADAPT-VQE further involves a global optimisation step requiring additional energy evaluations, this choice of shot-number results in GGA-VQE consuming a strictly fewer number of shots than its ADAPT-VQE counterpart. No Pauli grouping strategies were used in the measurement process. All simulations used a pool of non spin-complemented restricted single- and double-qubit excitation (QEB) evolutions as introduced in Section \ref{sec:Pools}. Moreover, all algorithms were executed up to a maximum of 40 iterations since we feel that this is a very optimistic upper-bound on the number of iterations achievable in the near-future on NISQ devices. Finally, each noisy simulation was performed five times, with the average energy value plotted and error bars representing the standard deviation across the runs.

Figure \ref{fig:GGA_1} displays the results of the GGA-VQE and ADAPT-VQE algorthms for computing the ground state energies of an H$_2$O and LiH molecule at equilibrium geometries. These energy plots support our conjecture that the local optimisation step in the GGA-VQE procedure is more resilient to measurement noise than the global optimisation step in ADAPT-VQE algorithm. Thus, in the case of the H$_2$O molecule, GGA-VQE is nearly twice as accurate as ADAPT-VQE after about 30 iterations. For the LiH molcule, we achieve even better results as the GGA-VQE is nearly five times more accurate than ADAPT-VQE. These results indicate that sacrificing the global optimisation step in adapative variational quantum algorithms in favour of local optimisation is indeed a useful strategy if we wish to counter measurement noise on quantum devices.

\begin{figure}[H]
  \centering
    \includegraphics[width=0.6\textwidth]{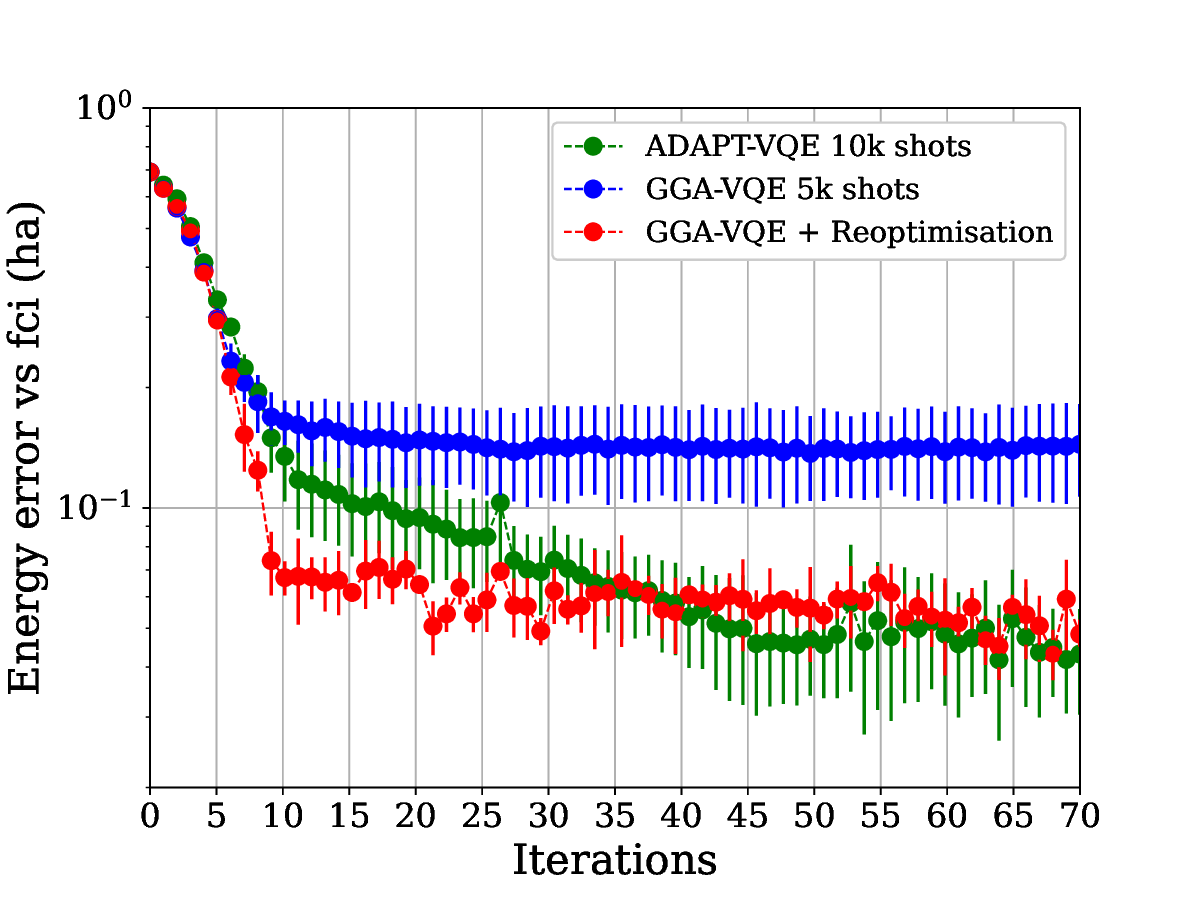}
  \caption{Comparison of the basic GGA-VQE algorithm, the \emph{sequentially reoptimised} GGA-VQE algorithm (in the style of RotoSolve \cite{ostaszewski2021structure}; see Algorithm \ref{alg:1} in the Appendix), and the ADAPT-VQE algorithm for the ground state energy of a stretched linear H$_6$ chain with bond distance of 3 Angstrom. The plots represents the energy convergence as a function of the number of iterations of the algorithms.. The simulation involved 12 qubits.}
  \label{fig:GGA_2}
\end{figure}

As a more stringent test, we next use GGA-VQE to compute the ground state of a stretched linear H$_6$ chain with a bond distance of $3$ Angstrom. The stretched H$_6$ chain is considered a benchmark test case since it is a prototypical example of a strongly correlated system. As emphasised in the introduction, tackling such strongly correlated systems is not the purpose of the present study but it is nevertheless interesting to evaluate the performance of the GGA-VQE algorithm. Our results are displayed in Figure \ref{fig:GGA_2}.

As expected for such a difficult system, both the ADAPT-VQE and GGA-VQE algorithms struggle to compute the ground state energy of the stretched H$_6$ chain. In the case of ADAPT-VQE, this is due to measurement noise since the noiseless implementation (not shown here; c.f. \cite{yordanov2021qubit}), while stagnating above chemical accuracy at around $3$ milliHartree, is an order of magnitude more accurate. The GGA-VQE computation is also far from accurate, having converged quickly to a shallow local minimum well-above the ground state. In our opinion, this rapid convergence indicates that for strongly correlated systems, a reoptimisation step is necessary to escape shallow minima. To test this conjecture, we supplement the basic GGA-VQE algorithm with a post-processed version in line with the discussion in Section \ref{sec:d-dimen}. More precisely, we plot the results for an improved version of GGA-VQE wherein, after each step of adding a locally optimised operator to the current GGA ansatz, we additionally perform a sequential reoptimisation of all previously chosen operators in the spirit of RotoSolve \cite{ostaszewski2021structure} as outlined in Algorithm \ref{alg:1} in Section \ref{sec:d-dimen} of the Appendix. As Figure \ref{fig:GGA_2} indicates, the post-processed GGA-VQE algorithm yields essentially the same accuracy after a dozen iterations as the ADAPT-VQE algorithm does after nearly 30 iterations. This result suggests that even for strongly correlated systems, a sequentially reoptimised GGA-VQE algorithm is competitive with ADAPT-VQE if measurement noise is present.

\subsection{The GGA-VQE algorithm applied to the Ising Model on a Quantum Device}

For our next set of numerical experiments, we turn to an actual realisation of the GGA-VQE algorithm on a physical quantum device. More specifically, we apply the GGA-VQE algorithm described in Section \ref{sec:Energy_Sorted} to the transverse-field Ising model described in Section \ref{sec:Energy_Sorted_Ising} on an IonQ Aria 25-qubit trapped-ion quantum computer. We set the system parameters of the Ising Hamiltonian to $h=0.5$, $J=0.2$, which ensures that the two-body interactions in this Ising Hamiltonian play an important role.

\begin{figure}[H]
    \centering
    \includegraphics[width=0.8\textwidth]{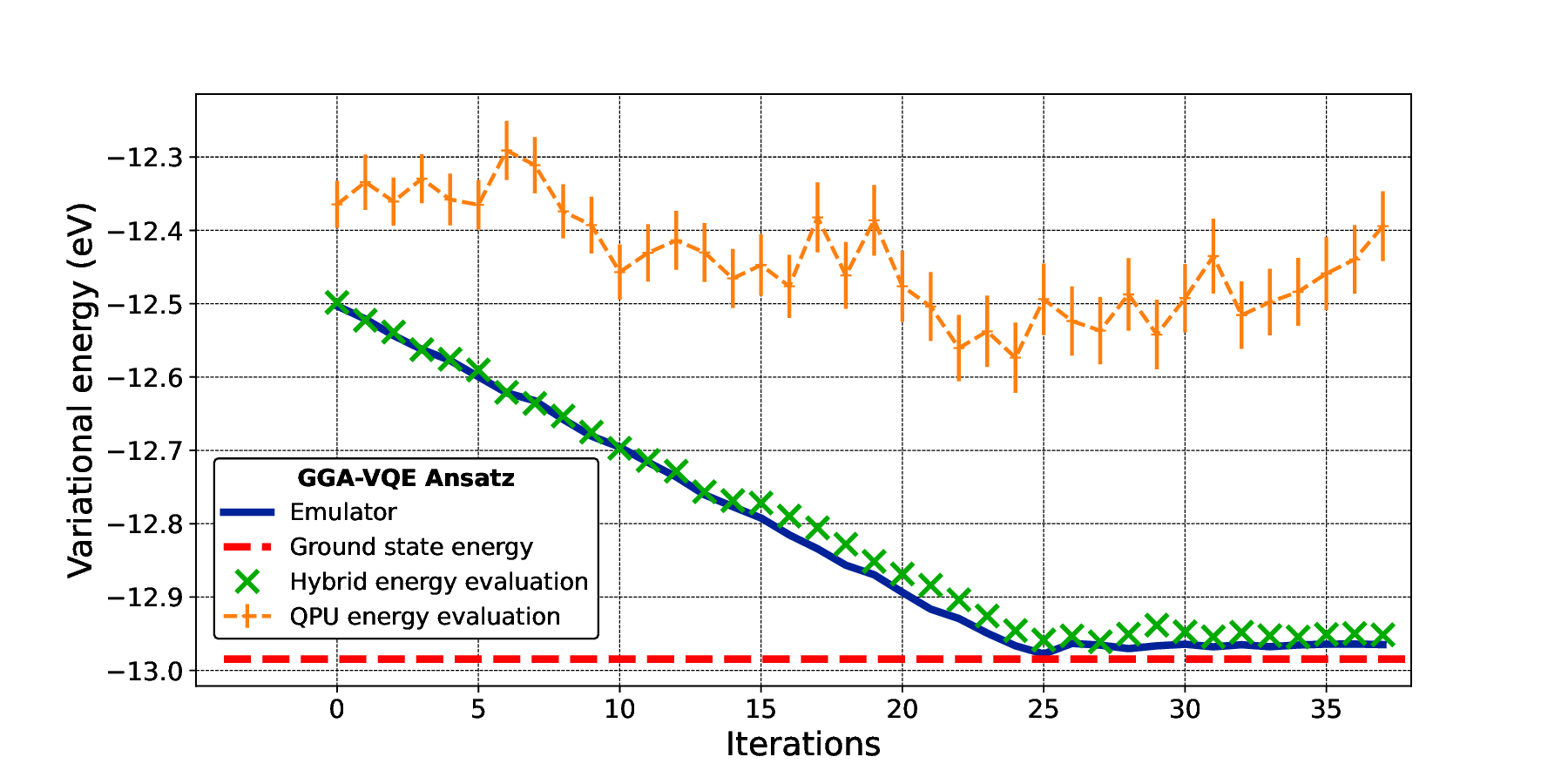}
    \caption{Energy convergence of the GGA-VQE algorithm with respect to the number of iterations. The blue reference curve denotes the energy of a classically simulated ansatz. The green and orange curves denote the hybrid and 25-qubit QPU energy evaluations of the GGA-VQE ansatz wave-function respectively. The hybrid evaluation is carried out by retrieving the GGA-VQE ansatz wave-function generated by the QPU, re-implementing it on the Hyperion-1 HPC emulator, and then evaluating the variational energy.}
    \label{fig:energy_convergence}
\end{figure}

\begin{figure}[H]
  \centering
  \begin{subfigure}[b]{0.45\textwidth}
    \centering
    \includegraphics[width=\textwidth]{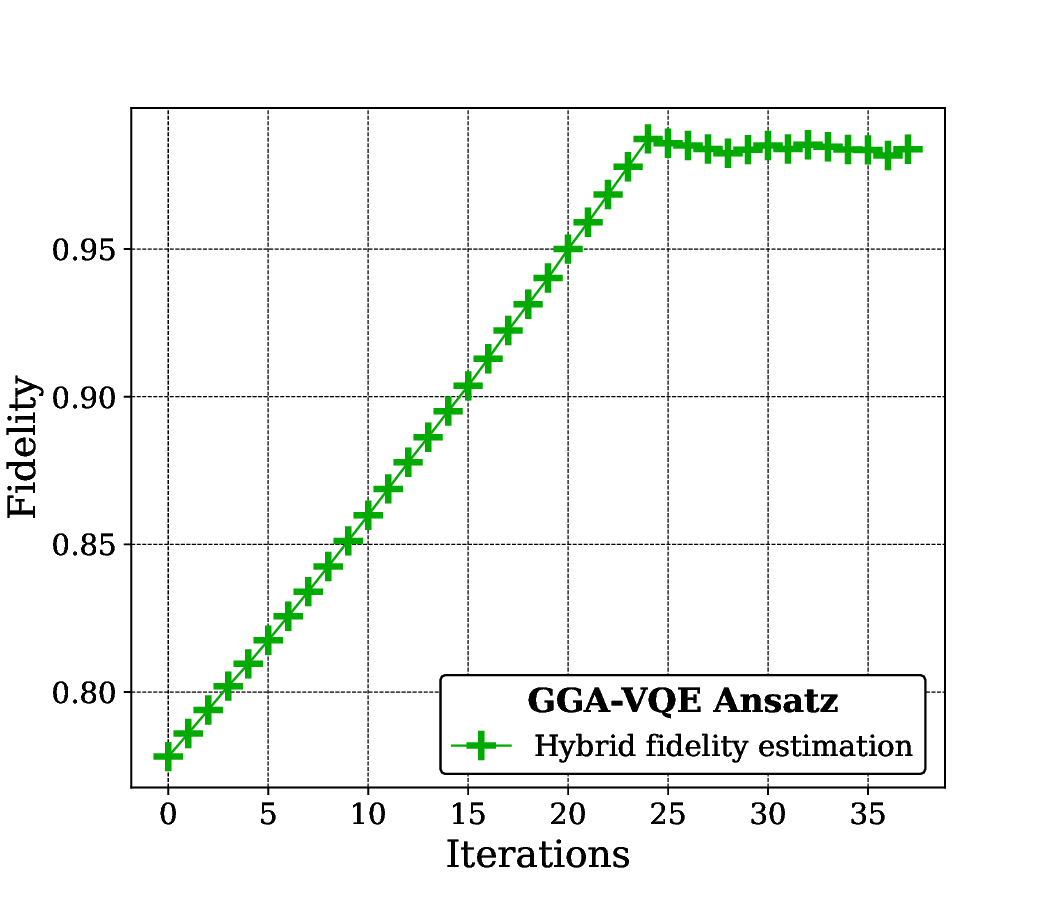}
  \end{subfigure}
    \begin{subfigure}[b]{0.45\textwidth}
    \centering
    \includegraphics[width=\textwidth]{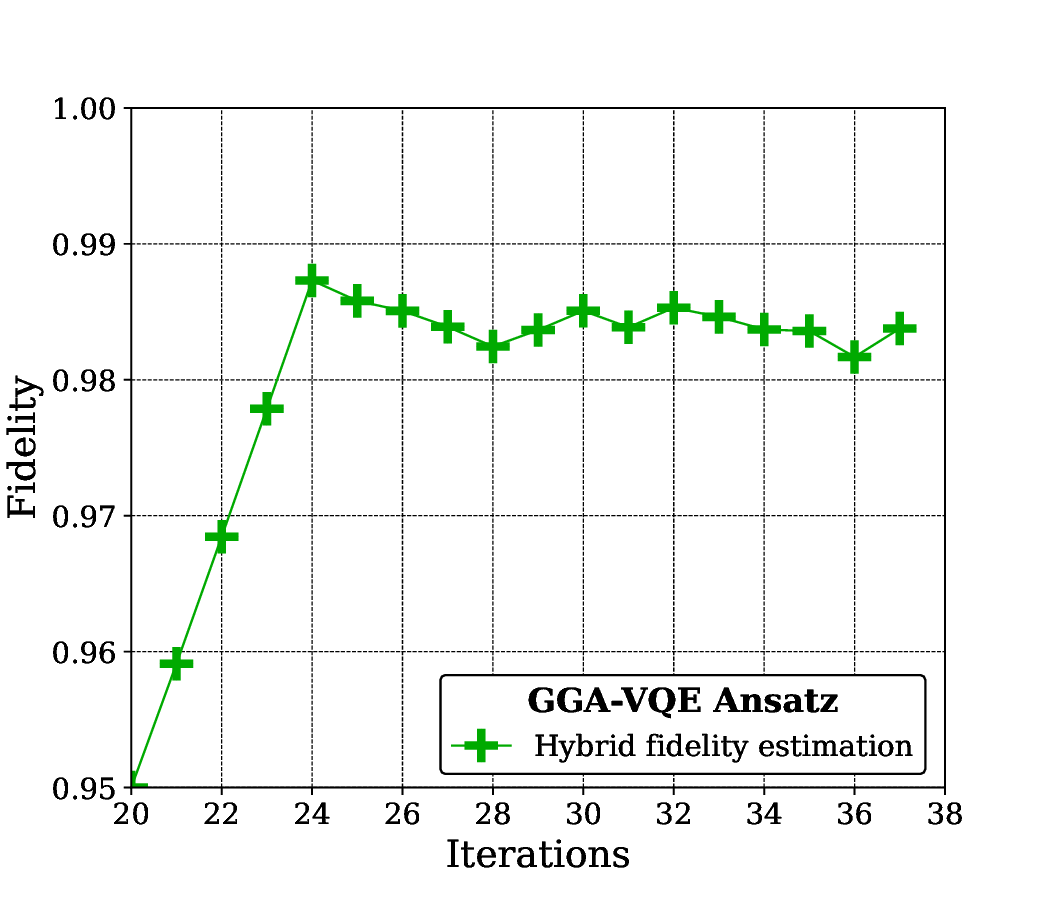}
  \end{subfigure}
  \caption{A convergence plot of the fidelity of the GGA-VQE ansatz wave-function produced by the 25-qubit QPU and re-implemented in the Hyperion-1 HPC emulator (hybrid evaluation approach) with the exact ground state of this Ising model obtained using a diagonalisation procedure on the Hyperion-1 emulator. The figure on the right is a zoomed-in version of the figure on the left to better appreciate the fidelity of the GGA-VQE ansatz wave-function. We herein refer to the fidelity of two quantum states $\ket{\psi}$ and $\ket{\phi}$ as the overlap squared of these two states, i.e., $F(\ket{\psi}, \ket{\phi}) = |\langle{\psi}{\ket{\phi}}|^2$.}
  \label{fig:ising_plot}
\end{figure}

Figure \ref{fig:energy_convergence} illustrates the convergence of the hybrid energy evaluations of the GGA-VQE ansatz wave-function with respect to the number of algorithm iterations. We remind the reader that these hybrid energy evaluations are obtained by first running the GGA-VQE algorithm on the IonQ Aria QPU, retrieving the resulting ansatz wave-function and re-implementing it on the Hyperion-1 HPC emulator, and then evaluating the variational energy on the HPC emulator. For reference, we also plot the corresponding energy curve obtained by executing the GGA-VQE ansatz directly on the HPC Hyperion-1 emulator using $10^6$ samples per measured circuit. In addition, as an indication of the measurement and hardware noise, we also plot the GGA-VQE energies obtain by direct measurement on the QPU.

Figure \ref{fig:ising_plot} clearly indicates that the QPU-implemented GGA-VQE procedure successfully provides an ansatz wave-function that closely matches the ground state. Moreover, the QPU implementation and HPC emulator implementation of the GGA-VQE algorithm seem highly consistent \emph{despite} significant noise in the quantum evaluation of observables, as noticeable from the QPU energy evaluation curve in Figure \ref{fig:energy_convergence}. Indeed, the greedy, gradient-free operator selection procedure that we have introduced in this study, which relies on a function extrapolation using five noisy evaluations on the QPU, is able to build an ansatz with an energy error below $2.50\times 10^{-2}$ eV and a fidelity exceeding 98\% with the exact ground state (see Figure \ref{fig:ising_plot}).

\begin{figure}[H]
  \centering
  \includegraphics[width=0.85\textwidth]{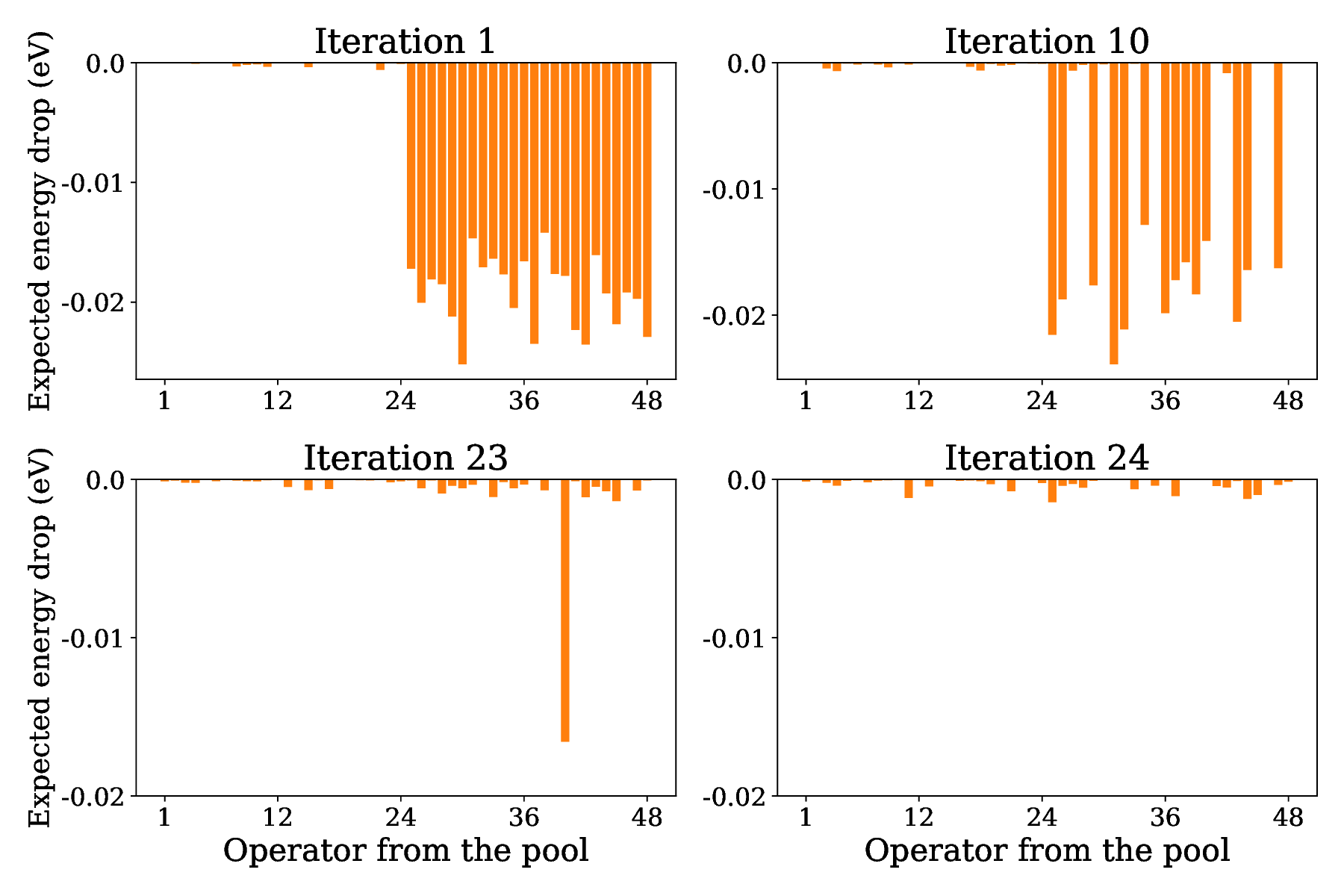}
  \caption{Expected energy drop of each Hermitian generator from the operator pool during the GGA-VQE iterative procedure on the QPU. The minimal pool operators, numbered from 1 to 48, are listed in the same order as defined in Equation \eqref{eq:Ising_Pool}. All energy differences are expressed in eV and all experiments involved 25 qubits.}
  \label{fig:landscapes_flattening}
\end{figure}

To better illustrate the robustness of the GGA-VQE procedure with respect to QPU noise, we depict in Figure \ref{fig:landscapes_flattening}, the expected maximal energy drop of each Hermitian generator from the chosen minimal operator pool throughout the iterative procedure. We observe maximal energy drops of approximately $1.5\times 10^{-2}$ eV for the first 24 iterations followed by a great reduction in the potential energy drops from iteration 25 onwards. This is consistent with the energy curve displayed in Figure~\ref{fig:energy_convergence} which steadily decreases for the first 24 iterations and then reaches a plateau. It is important to note that while Figure \ref{fig:energy_convergence} displays a decrease in the hybrid evaluation of energy between the 24th and 25th iterations, in our opinion, this decrease cannot be attributed to the newly introduced operator. Rather, it is simply a fortuitous consequence of measurement noise. Indeed, the reference emulator curve does not demonstrate any such energy decrease after the 24th iteration.

A few remarks concerning the results of Figures \ref{fig:ising_plot}- \ref{fig:landscapes_flattening} are now in order. First, as can be seen from Figure \ref{fig:landscapes_flattening}, at the beginning of the iterative procedure, there are 24 possible operators that can be appended to the initial state in order to achieve a sizeable lowering of the energy. These operators are all of the form $Z_{i}Y_{i+1}$ (recall the definition of the minimal hardware pool in Section \ref{sec:Energy_Sorted_Ising}). Second, we have confirmed from additional noiseless simulations that these operators do not all yield energy reductions of precisely the same amount in the course of the iterative procedure; there are indeed differences albeit small. Third, our noiseless tests also confirm that when exponeniated operators of this form are appended to the ansatz wave-function, the optimal angle for each operator is not identical. Fourth, our noiseless tests indicate that the energy of the ansatz wave-function is not very sensitive to the precise \emph{sequence} in which these operators are appended, i.e. the precise order in which the $Z_{i}Y_{i+1}$ operators are picked is not crucial. Taken together, these observations suggest that the Ising model that we study is indeed highly structured (although not perfectly so) and should therefore not be difficult to solve using classical techniques. We nevertheless feel that this is an appropriate test since the purpose of our experiment is simply to assess the performance of the GGA-VQE algorithm on a physical 25-qubit quantum device when applied to a simple problem.

An interesting question that can be raised at this stage is whether the analytical optimisation step in GGA-VQE, which is based on one-dimensional landscape function extrapolation, is resistant to device and measurement (statistical) noise. To answer this question, we consider the following simple test. Let us denote by $B_m$ and $\theta_m$ the Hermitian generator and corresponding optimised angles obtained after the $m^{\rm th}$ iteration of GGA-VQE \underline{using the QPU}. In other words, the GGA-VQE ansatz after the $m^{\rm th}$ iteration will be of the form
\begin{align*}
   \vert \Psi^{(m)}\rangle = e^{-\imath \theta_m B_m} e^{-\imath \theta_{m-1} B_{m-1}} \ldots e^{-\imath \theta_1 B_1} \vert \Psi^{(0)}\rangle,
\end{align*}
where $\vert \Psi^{(0)}\rangle$ is the initial state, and both $B_k$ and $\theta_k$ have been obtained through landscape function extrapolation based on noisy evaluations from the QPU.

In this setting, let us now define, for any iteration $m$, the `true' optimal angle $\theta_m^{\mathrm{opt}}$ as
\begin{align}\label{eq:resise_1}
   { \theta_m^{\mathrm{opt}} := \underset{\theta \in [-\pi, \pi)}{\text{argmin}} \left \langle \Psi^{(0)}\vert  e^{\imath \theta_1 B_1}  \ldots e^{\imath \theta_{m-1} B_{m-1}} e^{\imath \theta B_m}  H  e^{-\imath \theta B_m} e^{-\imath \theta_{m-1} B_{m-1}} \ldots e^{-\imath \theta_1 B_1}  \vert \Psi^{(0)} \right \rangle.}
\end{align}
In other words, $\theta_m^{\mathrm{opt}}$ is the `true' optimal angle obtained at iteration $k$ assuming that all the  Hermitian generators $B_m, B_{m-1}, \ldots, B_m$ and previously optimsed angles $\theta_{m-1}, \theta_{m-2}, \ldots \theta_{1}$ are fixed and taken from the previous computations on the QPU. The difference $\vert \theta_m - \theta_m^{\mathrm{opt}}\vert$ therefore measures purely the QPU error in optimising the angle at iteration $k$ of the GGA-VQE algorithm {assuming that the choice of operators as well as previous angles are fixed.}

To obtain the `true' optimal angle $\theta_m^{\mathrm{opt}}$ at a given iteration $m$, we can use the operators $B_m, B_{m-1}, \ldots, B_1$ and the previous optimized angles $\theta_{m-1}, \theta_{m-2}, \ldots \theta_{1}$ obtained from the QPU implementations, and then solve the optimisation problem on a noiseless HPC emulator. Our results are plotted below in Figure \ref{fig:revise_1} and indicate that, despite the significant fluctuations in the QPU-measured energy of the ansatz wave-function (Figure \ref{fig:energy_convergence}), the angles obtained through analytical optimisation based on QPU measurements are close to optimal for the first two dozen iterations.

\begin{figure}[h]
\begin{center}
    \includegraphics[width=0.5\textwidth]{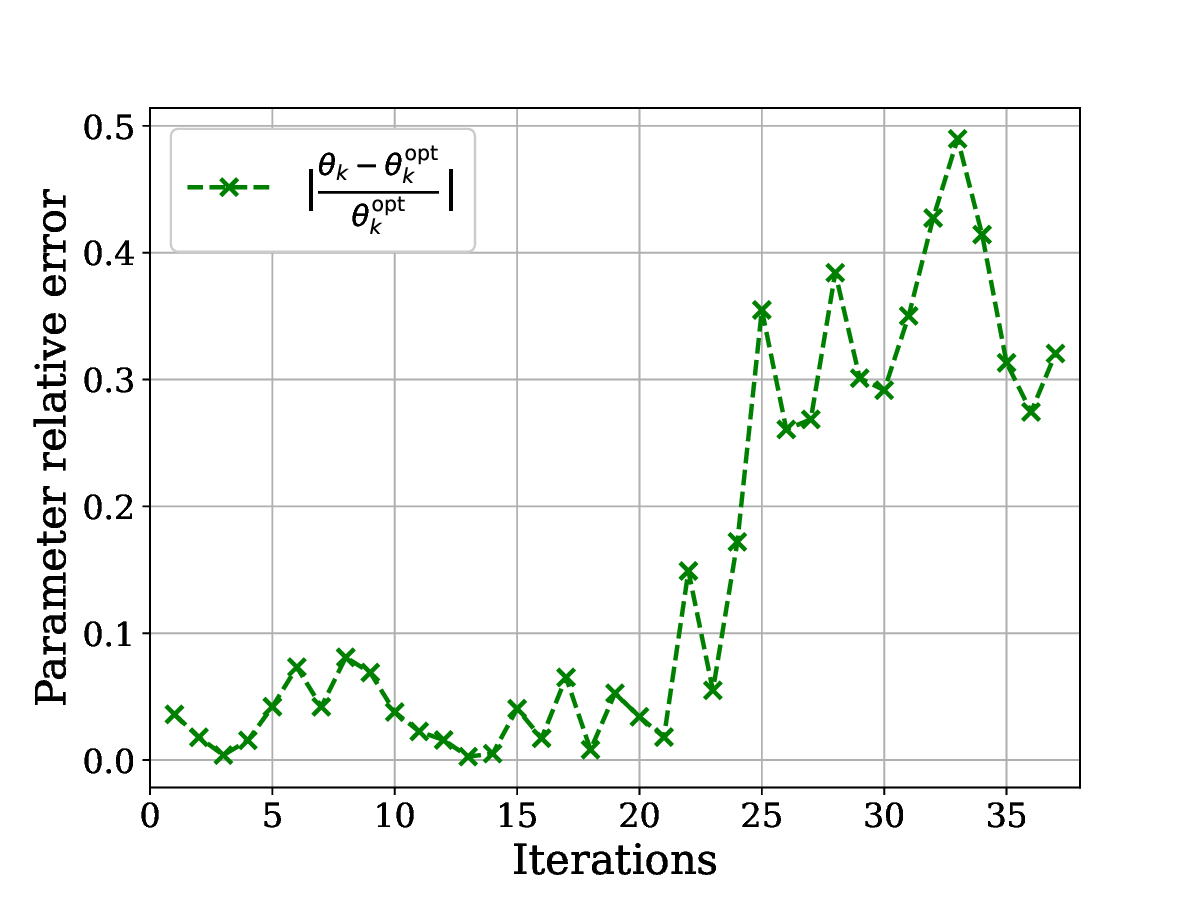}
\end{center}
    \caption{Relative error in optimising the angle at iteration $k$ of the GGA-VQE algorithm using only the QPU.}
    \label{fig:revise_1}
\end{figure}

\section{Discussion}\label{sec5}
In a recent study, Dalton et al. \cite{dalton2024quantifying} have quantified gate noise effects in ADAPT-VQE-type simulations, revealing that achieving a quantum advantage in a VQE algorithms would require a reduction in current hardware gate errors by several orders of magnitude. Even with ideal quantum gates, the successful execution of adaptive quantum algorithms requires tackling the challenge of measurement error due to statistical noise from observable sampling.  In light of these difficulties, we argue that noise-resistant strategies are needed to be able to successfully implement adaptive variational quantum algorithms on real quantum hardware. The two primary challenges that need to be addressed in this regard are the huge number of measurements required in the operator selection procedure in the ansatz growth stage and the subsequent multi-dimensional optimization of a highly noisy cost function. In the present study, we focus on the latter problem and introduce a new noise-resistant and resource-efficient, greedy gradient-free variational quantum algorithm (GGA-VQE) that relies on operator-by-operator local optimizations using only a small number of measurements on the quantum device.

As a first test of the proposed GGA-VQE algorithm, we have computed the ground states of the H$_2$O and LiH molecules on an HPC emulator in the presence of statistical noise arising from the finite number of observable evaluations. In both cases we show that the local optimisation step is more resilient to measurement noise than a global optimisation procedure and GGA-VQE thus outperforms ADAPT-VQE. Our main conclusion from these tests is that, for accessible, dynamically  correlated molecules, replacing a global optimisation step with local operator selection and optimisation is a successful strategy to deal with statistical noise.

As a more stringent test where a global reoptimisation of the ansatz wave-function is likely indispensable, we have applied the GGA-VQE algorithm to compute the ground state of a strongly correlated stretched linear H$_6$ chain. While both the ADAPT-VQE algorithm and the GGA-VQE algorithm perform poorly for this challenging system-- the former because of measurement noise and the latter because all optimisations are strictly local-- we show that a subsequent post-processing of the GGA-VQE ansatz wave-function based on a sequential local optimisation sweep (in the style of RotoSolve \cite{ostaszewski2021structure}) substantially improves the result and yields a very favourable comparison with ADAPT-VQE.

In addition to these HPC emulator results for molecular systems, we have also implemented the GGA-VQE on real  quantum architecture. More specifically, we have used the GGA-VQE algorithm to successfully compute the ground state of an open boundary 25-qubit transverse-field Ising Hamiltonian, achieving a ground state fidelity of over 98\%. The GGA-VQE algorithm that we have developed for the Ising model is also highly scalable since each iteration of this method requires a fixed number of circuit measurements, regardless of the number of qubits or the size of the operator pool. Ising models have already been studied using various methods in quantum regimes that claim to surpass the memory capacity of classical computers \cite{gacon2023stochastic, kim2023evidence}, and these studies, in combination with ours, demonstrate promising results in the potential of useful quantum computation before the era of fault-tolerance. 

Our results for the Ising model demonstrate that, despite the high level of device noise in observable quantum measurements, the GGA-VQE procedure can select a sequence of unitary operators and corresponding optimal angles that can be used to construct an accurate approximation of the ground state. Our greedy operator selection relies on an extrapolation of the associated objective function using a minimal number of noisy quantum measurements, and this extrapolation technique seems resilient to device noise, as evidenced by the close alignment between the QPU-extrapolated objective function and the HPC emulator-extrapolated objective function. 

Let us emphasize that the energy sorting procedure for optimal operator selection that we have developed in this study is easily extendable to multi-operator selection and optimization at the cost of a higher number of measurements on the quantum device, and some preliminary ideas in this direction have been presented in Section \ref{sec:d-dimen}. Similarly, the extensions of the Ising model GGA-VQE algorithm that we have developed (see Section \ref{sec:general_spin}) can easily be applied to other spin-chain systems such as the Hubbard model. Further research in both these directions will be the subject of future work.

We conclude this discussion by returning to the conclusion of Dalton et al. \cite{dalton2024quantifying} that the hardware performance window for reaching a quantum advantage in quantum chemistry using VQEs might well lie below the fault-tolerant threshold (see, also, \cite{tilly2022variational}). If confirmed, in our opinion, the partially successful implementation of GGA-VQE on the quantum hardware of today  and noisy HPC emulators indicates the suitability of these algorithms for approximate state preparation that can be used as the basis of a more accurate Quantum Phase Estimation (QPE) procedure \cite{abrams1999quantum, aspuru2005simulated ,tilly2022variational, feniou2024sparse} to evaluate the ground state energy of a given Hamiltonian. Since the probability of success for QPE is directly proportional to the fidelity between the approximate eigenstate and the true eigenstate, accurate, adaptive hybrid algorithms can play an important role in the pre-processing step for quantum phase estimation. Independent of such a pre-processing application however, we can also point to certain interesting studies that have demonstrated potential applications of adaptive algorithms to dynamic simulation problems\cite{yao2021adaptive, gomes2021adaptive}.

\section{Methods}

\subsection{The Adaptive Derivative-Assembled Pseudo-Trotter Variational Quantum Eigensolver}\label{sec:Adapt}

The adaptive derivative-assembled pseudo-Trotter variational quantum eigensolver (ADAPT-VQE)\cite{grimsley2019adaptive} is a VQE-inspired algorithm designed to approximate the ground state wave-function and ground state energy of a given Hamiltonian. Unlike many other classical variational quantum eigensolvers such as the various flavours of trotterised unitary coupled cluster \cite{tilly2022variational, romero2018strategies, ryabinkin2018qubit, kandala2017hardware, lee2018generalized, mizukami2020orbital} however, ADAPT-VQE does not specify a fixed ansatz for the sought-after ground state at the beginning of the algorithm. Instead, ADAPT-VQE functions by first fixing a set of admissible Hermitian generators (the so-called operator pool). The ansatz wave-function is then grown iteratively by parametrically exponentiating a carefully selected Hermitian generator, appending this exponentiated generator to the previous ansatz wave-function, and then variationally tuning the new ansatz wave-function. Since the selection procedure is tailored to the specific Hamiltonian system under consideration (see below), one usually hopes to obtain a more compact ansatz than the one generated by non-adaptive VQEs while still retaining the practical advantages of the VQE for near-term quantum hardware.

\vspace{2mm}

The general workflow of the ADAPT-VQE algorithm is as follows. Given the qubit representation of an input Hamiltonian~$H$, a pool of admissible Hermitian generators~$\mathbb{P}$, and a stopping criterion:
\begin{enumerate}
	\item Boot the qubits to an initial state $\ket{\Psi^{(0)}}$ and define the pool of parameterised unitary operator $\mathbb{U}$ as
 \begin{align}\label{eq:Yvon_new}
    \mathbb{U}:= \left\{\mathcal{U}(\theta)={\rm exp}(\imath \theta B)\colon \quad \theta \in [-\pi, \pi), B \in \mathbb{B}\right\}.
 \end{align}
	
\item At iteration $m$, we have at hand, a parameterised ansatz wave-function $\ket{\Psi^{(m-1)}}:=\ket{\Psi^{(m-1)}(\theta_{m-1}', \ldots, \theta_{1}')}$ of the form 
\begin{align*}
    \ket{\Psi^{(m-1)}}:=\mathcal{U}(\theta_{m-1}')\ldots \mathcal{U}(\theta_{1}') \ket{\Psi^{(0)}}.
\end{align*}
We identify the Hermitian generator $B_m \in \mathbb{P}$ such that the action of the parameterised unitary operator $\mathcal{U}(\theta_m)=\exp(-\imath \theta_m B_m), ~\theta_m \in [-\pi, \pi)$ on the current ansatz $\ket{\Psi^{(m-1)}}$ is \emph{likely} to produce a new wave-function with the largest drop in energy. This identification is done by computing the gradient, at $\theta=0$ 
, of the expectation value of the Hamiltonian, i.e., 
\begin{align}\label{eq:new_1}
     B_m = \argmax_{B\in\mathbb P}  \left \vert\frac{\partial}{{\partial \theta}} \braket{\Psi^{(m-1)}|\exp(\imath \theta B)H\exp(-\imath \theta B)|\Psi^{(m-1)}}\big\vert_{\theta=0} \right \vert.
\end{align}

Note that the criterion \eqref{eq:new_1} is simply a heuristic, and there is no guarantee that the Hermitian generator $B_m$ selected through this criterion will indeed lead to the parameterised unitary operator whose action on the current ansatz $\ket{\Psi^{(m-1)}}$ results in the largest drop in energy. This point will be the subject of further discussion in Section \ref{sec:Energy_Sorted}.

\item Exit the iterative process if the stopping criterion is met (see below for more explanation). Otherwise, append the resulting parametrised unitary operator to the left of the current ansatz wave-function $\ket{\Psi^{(m-1)}}$, i.e., define 
\begin{align*}
    \ket{\widetilde{\Psi^{(m)}}}:=& \exp(-\imath \theta_mB_m)\ket{\Psi^{(m-1)}}
    =\exp(-\imath \theta_mB_m)\exp(-\imath \theta_{m-1}'B_{m-1})\ldots \exp(-\imath \theta_1'B_1)\ket{\Psi^{(0)}}.
\end{align*}

\item Run a classical VQE routine by optimising all parameters $\theta_m, \theta_{m-1}, \ldots, \theta_1$ in the new ansatz wave-function $ \ket{\widetilde{\Psi^{(m)}}}$ so as to minimize the expectation value of the Hamiltonian, i.e., solve the optimisation problem
    \begin{align}\label{eq:adapt_opt}
        \vec{\theta}^{\rm{opt}} :=& (\theta_1', \ldots, \theta_{m-1}', \theta_m')\\ \nonumber
        :=& \underset{\theta_1, \ldots, \theta_{m-1}, \theta_{m}}{\operatorname{argmin}}\Big \langle\Psi^{(0)}\Big| \prod_{k=1}^{k=m} \exp(\imath \theta_kB_k)H\prod_{k=m}^{k=1} \exp(-\imath \theta_kB_k)\Big \vert \Psi^{(0)} \Big \rangle,
    \end{align}
    and define the new ansatz wave-function $\ket{\Psi^{(m)}}$ using the newly optimized parameters $\theta_1', \ldots, \theta_m'$, i.e., define 
    \begin{align*}
    \ket{\Psi^{(m)}}:= \prod_{k=m}^{k=1}\exp(-\imath \theta_k'B_k)\ket{\Psi^{(0)}} .  
    \end{align*}
    
    Let us emphasize, as indicated in above, that although we also denote the newly optimized parameters at the current $m^{\rm th}$ iteration by $\theta_1',\ldots \theta_m'$, these optimized values are not necessarily the same as those used to define $\ket{\Psi^{(m-1)}}$ and referenced in Step 3 above.

    \item Return to Step 2 with the updated ansatz $\ket{\Psi^{(m)}}$. 

\end{enumerate}

\vspace{5mm}

Let us remark here that a common choice of stopping criterion is to impose a pre-defined threshold tolerance $\epsilon>0$ on the magnitude of the gradients computed in Step~2 above, i.e., exit the ADAPT-VQE algorithm at iteration $m$ if
    \begin{align*}
   \max_{B\in\mathbb P}  \left \vert\frac{\partial}{{\partial \theta}} \braket{\Psi^{(m-1)}|\exp(\imath \theta B)H\exp(-\imath \theta B)|\Psi^{(m-1)}}\big\vert_{\theta=0} \right \vert < \epsilon.
    \end{align*}
An obvious alternative option is to impose a maximal iteration count on the number of ADAPT-VQE steps or a minimal decrease of the expectation value between two iterates
\begin{equation*}
    \Big \langle\Psi^{(0)}\Big| \prod_{k=1}^{k=m-1} \exp(\imath \theta'_kB_k)H\prod_{k=m-1}^{k=1} \exp(-\imath \theta'_kB_k)\Big \vert \Psi^{(0)} \Big \rangle - \Big \langle\Psi^{(0)}\Big| \prod_{k=1}^{k=m} \exp(\imath \theta'_kB_k)H\prod_{k=m}^{k=1} \exp(-\imath \theta'_kB_k)\Big \vert \Psi^{(0)} \Big \rangle< \varepsilon
\end{equation*}

\vspace{3mm}

Next, let us discuss some commonly used operator pools for ADAPT-VQE.

\subsection{Operator Pools for ADAPT-VQE}\label{sec:Pools}\vspace{2mm}

As one may expect by studying the ADAPT-VQE workflow, the success of the algorithm is strongly impacted by the choice of the Hermitian generator pool $\mathbb P$. As an extreme example, if all generators in the operator pool commute with the Hamiltonian, then the algorithm will terminate at the first iteration thus resulting in no improvement of the initial guess. The goal of this section is to briefly present some popular operator pools. While a great variety of operator pools have been introduced in the literature \cite{wiersema2020exploring, kandala2017hardware, romero2022solving}, we will limit ourselves to a `chemically-inspired' pool which is popular for simulating quantum chemical systems \cite{yordanov2021qubit}, a simplified version of this chemically-inspired pool which (empirically) leads to lower quantum gate counts \cite{tang2021qubit}, and finally a so-called \emph{minimal} operator pool which possesses some useful mathematical properties. Before doing so however, let us first clarify some additional details concerning the ADAPT-VQE algorithm.

ADAPT-VQE was primarily developed for quantum chemistry applications, i.e., for application to molecular systems typically described by a second-quantized molecular Hamiltonian mapped to a qubit representation through the Jordan-Wigner transformation. In this setting, an obvious choice for the initial state in the ADAPT-VQE algorithm is the Hartree-Fock wave-function. Indeed, in the standard formalism where each qubit is used to represent a specific spin-orbital, we can write $\ket{0}_p$ and $\ket{1}_p$ to denote states corresponding to an \emph{empty} and \emph{occupied} spin-orbital $p$ respectively. With this notation, the reference Hartree-Fock state for a system having $n$ electrons in $N$ spin-orbitals can be expressed as $\ket{\Psi_{\text{\rm HF}}}:= \ket{1_{0}\ldots 1_{n-1}0_{n}\ldots 0_{N-1}}$, which is straightforward to represent on quantum circuits. Note that this is an example of a case where we have access to an initial state that is both simple to represent on quantum architecture and also yields a wave-function having a reasonable overlap with the sought-after ground state wave-function. Of course, for arbitrary Hamiltonians, such an efficient choice for the initial state might not be possible, in which case we might have to rely on random initialisations, for instance.

\vspace{2mm}

\noindent \textbf{The Qubit Excitation-based Pool \cite{yordanov2021qubit}}\vspace{2mm}

The first commonly used operator pool is the Qubit excitation-based (QEB) pool which is inspired by the popular coupled cluster method from computational quantum chemistry. The QEB pool consists of so-called single-qubit and double-qubit excitation operators which take the form

\begin{align}\label{eq:single}
A_{pq}&=\frac{1}{2}\left(X_qY_p-Y_qX_p\right), \quad \text{and}\\[0.25em] \label{eq:double}
 A_{pqrs}&=\frac{1}{8}\left(
X_rY_sX_pX_q+Y_rX_sX_pX_q+Y_rY_sY_pX_q+Y_rY_sX_pY_q -X_rX_sY_pX_q-X_rX_sX_pY_q-Y_rX_sY_pY_q-X_rY_sY_pY_q\right). 
\end{align}
Here $p, q, r, $ and $s$ denote qubit indices and $X_p$ and $Y_p$ are the usual one-qubit Pauli gates acting on qubit $p$. Thus, the single-qubit generator $A_{pq}$ acts between the single qubits $p$ and $q$ while the double-qubit generator $A_{pqrs}$ acts between the qubit pairs $(p, q)$ and $(r, s)$.

Given an $N$-qubit system, it is easy to see that the QEB operator pool a priori has $\mathcal{O}(N^4)$ elements. In practice however, not all possible excitation operators of the form \eqref{eq:single}-\eqref{eq:double} are included in the operator pool. Instead, the QEB operator pool is limited to those single-qubit and double-qubit excitation operators which preserve important symmetries in the system such as spin or the number of particles. Additionally, it is readily checked that the parametric exponentiation of a QEB operator is easy to calculate, and the resulting unitary operators have well-known CNOT-optimised circuits (see, e.g., Figure \ref{fig:d_q_exc_full}
in the appendix).

\vspace{5mm}

\noindent \textbf{The Qubit Hardware-efficient Pool \cite{tang2021qubit}}\vspace{2mm}

While the Qubit excitation-based pool provides excellent performance in numerical simulations on quantum emulators, the practical implementation of QEB-based ansatz wave-functions on near-term quantum hardware remains challenging. This is primarily due to the fact that the number of CNOT gates required to construct the associated QEB circuits, while significantly smaller than the CNOT counts for classical ``fixed-ansatz'' approaches, is still far too high. The so-called Qubit hardware-efficient pool \cite{ryabinkin2018qubit} addresses this issue by considering instead a pool consisting of \emph{modified} single and double excitation operators of the form
\begin{align}\label{eq:single_2}
\widetilde{X}_{pq}=X_qY_p
\end{align}
and
\begin{equation}\label{eq:double_2}
\begin{split}
 \widetilde{X}^{(1)}_{pqrs}=&X_rY_sX_pX_q, \quad \widetilde{X}^{(2)}_{pqrs}= Y_rX_sX_pX_q, \quad \widetilde{X}^{(3)}_{pqrs}= Y_rY_sY_pX_q, \quad \widetilde{X}^{(4)}_{pqrs}=Y_rY_sX_pY_q,\\ 
 \widetilde{X}^{(5)}_{pqrs}=& X_rX_sY_pX_q, \quad \widetilde{X}^{(6)}_{pqrs}=X_rX_sX_pY_q, \quad \widetilde{X}^{(7)}_{pqrs}= Y_rX_sY_pY_q, \quad \widetilde{X}^{(8)}_{pqrs}= X_rY_sY_pY_q,
\end{split}
\end{equation}
where $p, q, r, s$ again denote qubit indices and $X_p$ and $Y_p$ are one-qubit Pauli gates acting on qubit $p$. Note that not all \emph{modified} double qubit-excitation operators of the form \eqref{eq:double_2} are added to the operator pool since, for instance, the operator $\widetilde{X}^{(1)}_{pqrs}=X_rY_sX_pX_q$ and $\widetilde{X}^{(7)}_{pqrs}=Y_rX_sY_pY_q$ are related by a global rotation \cite{tang2021qubit}. Let us also emphasize here that the qubit hardware-efficient pool is \emph{not particle conserving} and thus violates an important symmetry of the system.

Numerical experiments involving the qubit hardware-efficient pool demonstrate that the CNOT count of the resulting ansatz wave-functions is very competitive with the CNOT count of the QEB pool-based ansätze.
\vspace{5mm}

\noindent \textbf{Minimal Hardware-efficient Pool \cite{tang2021qubit}}\vspace{2mm}

It is well-known that when the Hamiltonian under study is real-valued, then the ground state eigenfunction can be expressed as a \emph{real} linear combination of \emph{real} basis vectors. For such settings therefore, it is possible to show the existence of a so-called minimal operator pool that allows the transformation of any real-valued wave-function (in particular, the Hartree-Fock reference state) to another real-valued wave-function (in particular, the sought-after ground state eigenfunction). More precisely, given an $N$-qubit system, we may define the operator pool 
\begin{align*}
\mathbb P=\{Y_p\}_{p=0}^{N-2} ~\cup ~\{Z_pY_{p+1}\}_{p=0}^{N-2},
\end{align*}
where $Y_p$ and $Z_p$ are the usual one-qubit Pauli gates acting on qubit $p$. Then, for any two real-valued wave-functions $\ket\Phi$ and $\ket\Psi$, there exist $\theta_1, \ldots, \theta_M \in [-\pi, \pi)$ and Hermitian generators $B_1, \ldots, B_M \in \mathbb{P}$ such that
\begin{align*}
\ket\Phi=\prod_{k=M}^{1}\exp(-\imath\theta_k B_k)\ket\Psi.
\end{align*}
In other words, for a precise choice of parameters and generators in the pool $\mathbb{P}$, we can construct the sought-after ground-state eigenfunction by applying the parametrised, exponentiated generators to the Hartree-Fock reference state.

The key advantage of this so-called minimal hardware-efficient pool $\mathbb P$ is that it consists of only $2N-2$ elements, and the operators in this pool can be parametrically exponentiated using very simple circuits that require a minimal number of CNOT gates.

\vspace{5mm}

\subsection{Resource Saving Enhancements and the GGA-VQE Algorithm}\label{sec:Energy_Sorted}

\begin{figure}[H]
    \centering
    \includegraphics[width=0.5\textwidth]{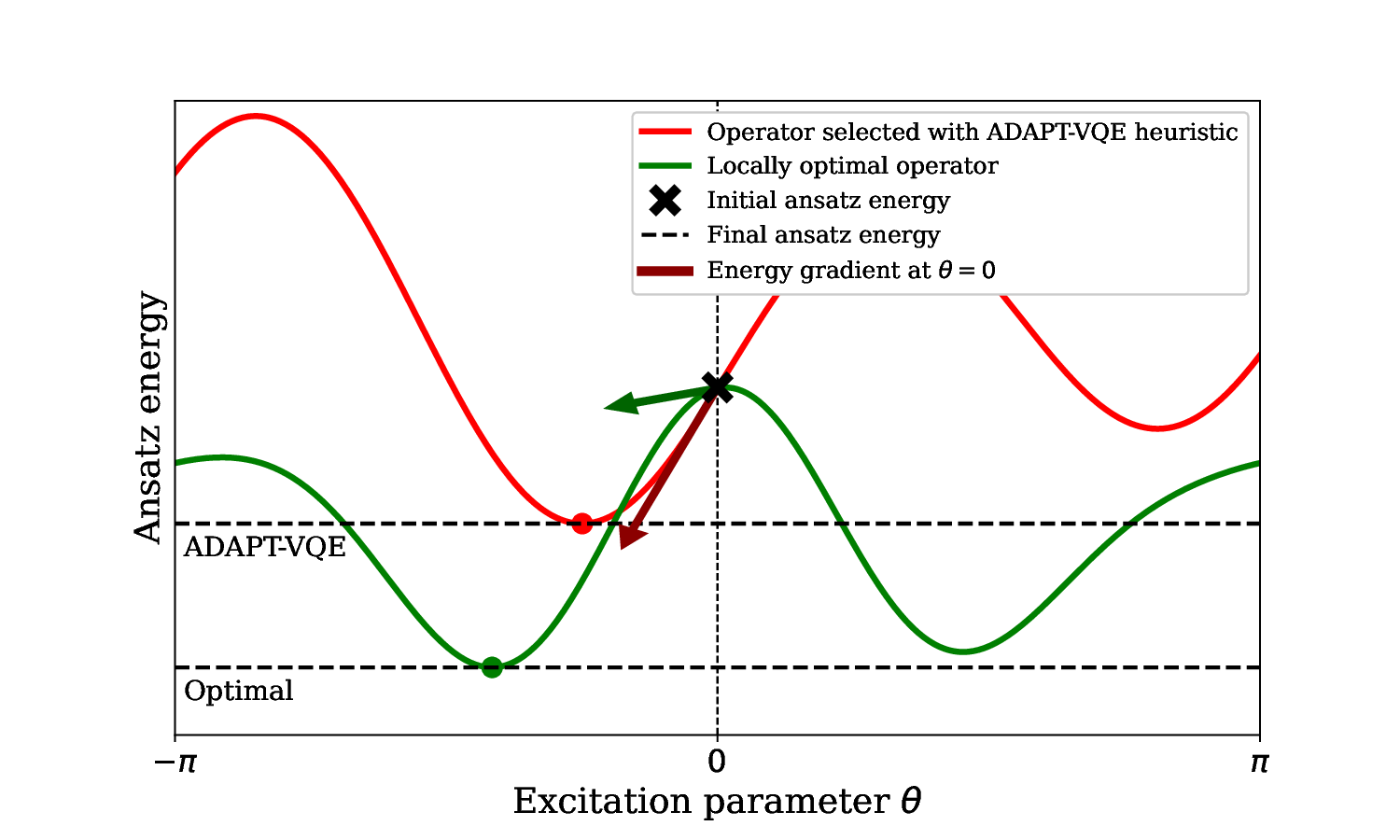}
    \caption{Illustration of a situation where the ADAPT-VQE selection criterion does not pick the optimal operator leading to the largest energy drop.}
    \label{fig:adapt_heuristic}
\end{figure}

As described in detail in Section \ref{sec:Adapt}, a core step in the ADAPT-VQE algorithm is the selection of an optimal Hermitian generator $B$ from the operator pool whose addition to the current ansatz can produce a new ansatz wave-function with the largest drop in energy. Current implementations of ADAPT-VQE make this choice through a heuristic criterion based on evaluating certain gradients of the expectation value of the Hamiltonian. More precisely, for a given pool of operators $\mathbb{P}$, at the $m^{\rm th}$ iteration, one computes (c.f., Equation \eqref{eq:new_1})
\begin{align}\label{eq:new_11}
    B_m = \underset{B\in\mathbb P}{\text{\rm argmax}}  \left \vert\frac{\partial}{{\partial \theta}} \braket{\Psi^{(m-1)}|\exp(\imath \theta B)H\exp(-\imath \theta B)|\Psi^{(m-1)}}\big\vert_{\theta=0} \right \vert,
\end{align}

where $\ket{\Psi^{(m-1)}}$ denotes the ansatz wave-function at the ${m-1}$ iteration. Of course, Equation \eqref{eq:new_11} is still a heuristic, and it is conceivable that there exists another Hermitian generator $\widetilde{B}_m$ such that 
\begin{align*}
    \min_{\theta \in [-\pi, \pi)}  \braket{\Psi^{(m-1)}|\exp(\imath \theta \widetilde{B}_m)H\exp(-\imath \theta \widetilde{B}_m)|\Psi^{(m-1)}}&<\min_{\theta \in [-\pi, \pi)}  \braket{\Psi^{(m-1)}|\exp(\imath \theta {B}_m)H\exp(-\imath \theta {B_m})|\Psi^{(m-1)}}\\
    \intertext{while}
  \left \vert\frac{\partial}{{\partial \theta}} \braket{\Psi^{(m-1)}|\exp(\imath \theta \widetilde{B}_m)H\exp(-\imath \theta \widetilde{B}_m)|\Psi^{(m-1)}}\big\vert_{\theta=0} \right \vert &< \left \vert\frac{\partial}{{\partial \theta}} \braket{\Psi^{(m-1)}|\exp(\imath \theta {B}_m)H\exp(-\imath \theta {B}_m)|\Psi^{(m-1)}}\big\vert_{\theta=0} \right \vert.
\end{align*}
A representative example of this situation is displayed in Figure \ref{fig:adapt_heuristic}.

Given that at each iteration of ADAPT-VQE, we face the task of optimising a multi-dimensional objective function which is very noisy due to the quality of the current quantum hardware, the selection of the wrong operator to append to the current ansatz wave-function can be a costly mistake \cite{grimsley2023adaptive}. In this section, we introduce an energy sorting algorithm that allows the exact selection of the \emph{locally, optimal} Hermitian generator from any of the three operator pools that we have introduced in Section~\ref{sec:Pools}. In other words, we show that it is possible, using a few evaluations of Hamiltonian expectations on the quantum device, to exactly solve the optimisation problem
\begin{align}\label{eq:new_opt}
    B_m= \underset{B \in \mathbb{P}}{\text{argmin}}  \min_{\theta \in [-\pi, \pi)} \mathcal{L}(B, \theta, \ket{\Psi^{(m-1)}}) :=  \underset{B \in \mathbb{P}}{\text{argmin}} \min_{\theta \in [-\pi, \pi)}\braket{\Psi^{(m-1)}|\exp(\imath \theta B)H\exp(-\imath \theta B)|\Psi^{(m-1)}},
\end{align}
where, for notational convenience, we have introduced the objective function $\mathcal{L}(B, \theta, \ket{\Psi^{(m-1)}})$, which, in some literature, is also referred to as a \emph{landscape} function. \vspace{2mm}

The main idea of our energy sorting algorithm is to take advantage of a result that is largely known in the VQE literature, namely that the expectation value of a Hermitian operator with respect to an ansatz wave-function $\ket{\Psi(\theta)}$ parametrised by a single quantum gate can be expressed in terms of elementary trigonometric functions of $\theta$ \cite{watanabe2023optimizing, ostaszewski2021structure, wada2022sequential, liu2023training, nakanishi2020sequential}. While this result is typically expressed for ansatz wave-function parametrised by rotation gates, the core arguments can be extended to wave-functions parametrised by Hermitian generators belonging to any of the three operator pools that we have introduced in Section \ref{sec:Pools}. Indeed, taking advantage of the precise functional form of the Hermitian generators introduced in Section~\ref{sec:Pools}, a simple calculation shows that
\begin{enumerate}
    \item For any generator $B$ in the Qubit-Excitation-Based (QEB) operator pool, it holds that $B^3=B$ with $I$ denoting the identity matrix (see Appendix \ref{appendix_pools}).

    \item For any generator $B$ in the Qubit hardware-efficient and minimal hardware-efficient pools, it holds that $B^2=I$ .
\end{enumerate}

The above simple relations now imply (see also Appendix \ref{appendix_pools}) that for any generator $B$ in the Qubit-Excitation-Based (QEB) operator pool and any $\theta \in [-\pi, \pi)$, it holds that 
    \begin{align}\label{eq:new_has_1}
        \exp(-\imath \theta B)=I+(\cos(\theta)-1)B^2-\imath\sin(\theta) B,
    \end{align}
and for any generator $B$ in the Qubit hardware-efficient and minimal hardware-efficient pools and any $\theta \in [-\pi, \pi)$, it holds that 
    \begin{align}\label{eq:new_has_2}
        \exp(-\imath \theta B)=\cos(\theta)I-\imath\sin(\theta) B.
    \end{align}

Using Equations \eqref{eq:new_has_1} and \eqref{eq:new_has_2}, it is easy to establish that for any wave-function $\ket{\phi}$ and any $\theta \in [-\pi, \pi)$ the objective function $\mathcal{L}(B, \theta, \ket{\phi})$ has the analytical form
 \begin{equation}\label{eq:landscape}
\mathcal{L}(B, \theta, \ket{\phi}) =\begin{cases} \braket{\phi|H|\phi}
+\big(\cos(\theta)-1\big)(\braket{\phi|\{H,B^2\}|\phi}-2\braket{\phi|BHB|\phi} ) \hspace{2cm} \text{if } B^3=B,\\[0.5em]
+\big(1-\cos(\theta)\big)^2\big(\braket{\phi|B^2HB^2|\phi}-\braket{\phi|BHB|\phi} \big)\\[0.5em]
+\sin(\theta)(\cos(\theta)-1)\braket{\phi|\imath B[H,B]B|\phi} \\[0.5em]
+\sin(\theta)\braket{\phi|\imath[B,H]|\phi}\\[2em]
\cos^2(\theta)\braket{\phi|H|\phi} +\dfrac{\sin(2\theta)}{2}\braket{\phi|\imath [B,H]|\phi} \hspace{4.15cm}\text{if } B^2=I.\\[0.75em]
+\sin^2(\theta)\braket{\phi|BHB|\phi},
\end{cases}
\end{equation}
where $\{\cdot, \cdot\}$ and $[\cdot, \cdot]$ denote the anti-commutator and commutator respectively. A demonstration of this result can be found in Appendix \ref{appendix_pools}. 

Equation \eqref{eq:landscape} implies that for any Hermitian generator $B$ from our operator pools and any arbitrary wave-function $\ket{\phi}$, the objective function $\mathcal{L}(B, \theta, \ket{\phi})$ can be expressed in terms of elementary trigonometric functions of $\theta$. An important consequence of this expression is that, if we now evaluate the objective function $\mathcal{L}(B, \theta, \ket{\phi})$ at certain well-chosen angles, we can obtain a linear system of equations for the unknown operator expectation values. More precisely,

\begin{description}
    \item[\rm \textbf{For the QEB pool ($B^3=B$):}]~ 
    
    Since $\braket{\phi|H|\phi}$ can be measured directly on the quantum device, four measurements of $\mathcal{L}(B, \theta, \ket{\phi})$ at well-chosen $\theta= \theta^{(1)}, \theta^{(2)}, \theta^{(3)}, \theta^{(4)}$ yields a linear system for the four unknowns $\braket{\phi|\{H,B^2\}|\phi}-2\braket{\phi|BHB|\phi}$, $\braket{\phi|B^2HB^2|\phi}-\braket{\phi|BHB|\phi}$, $\braket{\phi|iB[H,B]B|\phi}$ and $\braket{\phi|\imath[B,H]|\phi}$.\vspace{2mm}

    \item[\rm \textbf{For the hardware efficient pools ($B^2=I$):}]~
    
    Since $\braket{\phi|H|\phi}$ can be measured directly on the quantum device, two measurements of $\mathcal{L}(B, \theta, \ket{\phi})$ at well-chosen $\theta= \theta^{(1)}, \theta^{(2)}$ yields a linear system for the two unknowns $\braket{\phi|i[B,H]|\phi}$ and $\braket{\phi|BHB|\phi}$.
\end{description}

In other words, using a minimal number of evaluations of Hamiltonian expectations and by solving a very small linear system, we can compute all terms involving $B, H$, and $\ket{\phi}$ in the expression \eqref{eq:landscape} for the objective function $\mathcal{L}(B, \theta, \ket{\phi})$. Since the dependency of this function on $\theta$ is through elementary trigonometric functions, we can thus express the objective function $\mathcal{L}(B, \theta, \ket{\phi})$ analytically for any generator $B$, any angle $\theta$, and any wave-function $\ket{\phi}$. 
This allows us to solve the optimisation problem \eqref{eq:new_opt} up to arbitrary precision for any Hermitian generator $B$ from our operator pool, and thereby obtain the \emph{locally, optimal} generator that should be added to the current ansatz~wave-function $\ket{\Psi^{(m-1)}}$. It is pertinent to also mention here the article \cite{endo2023optimal} which studies methods to evaluate the landscape function $\mathcal{L}$ so as to minimize statistical noise uncertainty. Note that if we assume an operator pool of size $M$, a total of $4M+1$ evaluations of Hamiltonian expectations will be required to screen all Hermitian generators from the qubit-excitation based pool while a total of $2M+1$ Hamiltonian evaluations will be required to screen all Hermitian generators from the two hardware efficient pools.

Two important remarks are now in order. First, as mentioned previously, analytical expressions such as \eqref{eq:landscape} for landscape functions $\mathcal{L}(B, \theta, \ket{\phi})$ of the form \eqref{eq:new_opt} are present in the existing literature \cite{nakanishi2020sequential, ostaszewski2021structure, Wada_2022, watanabe2023optimizing, crooks2019gradients, izmaylov2021analytic}, albeit not- to the best of our knowledge- for the specific operator pools that we have considered in the current study. However, these landscape functions are almost exclusively used to perform an operator-by-operator, analytical optimisation of a structurally fixed, parameterised quantum circuit. Indeed, the only numerical method we are aware of that extends the applicability of analytical landscapes functions beyond simple iterative optimisation of a fixed ansatz circuit is the Rotoselect algorithm \cite{ostaszewski2021structure}, and Rotoselect still assumes a fixed structure for the parametrised quantum circuit, in which only the choice of $X, Y,$ or $Z$ rotation gate (and not the `location') can be varied according to the associated landscape function.

Second, let us point out that the landscape function \eqref{eq:new_opt} assumes the addition of a single operator to the current ansatz wave-function at each iteration of the adaptive algorithm. If the pool of potential unitary operators is commutative, then the specific order in which operators are chosen is unimportant, and it is therefore sufficient to consider a sequential application of the representation \eqref{eq:landscape} of $\mathcal{L}(B, \theta, \ket{\phi})$ to determine, at each iteration, the optimal operator to append to the current ansatz. On the other hand, if the Hermitian generators belonging to the operator pool do not commute (which is often the case), then the ordering of the operators is important, and it is potentially useful to consider landscape functions based on the simultaneous addition of $d>1$ operators to the current ansatz wave-function at each iteration. We now briefly discuss this generalisation.\vspace{5mm}

\noindent \textbf{The Greedy Gradient-free Adaptive Variational Quantum Eigensolver (GGA-VQE)}\vspace{1mm}

Given the qubit representation of an input Hamiltonian $H$, a pool of admissible Hermitian generators~$\mathbb{P}$, and a stopping criterion:
\begin{enumerate}
	\item Boot the qubits to an initial state $\ket{\Psi^{(0)}}$. \vspace{2mm}
	
\item At the $m^{\rm{th}}$ iteration, use the energy sorting algorithm detailed above to identify the Hermitian generator $B_m \in \mathbb{P}$ that solves the optimisation problem \eqref{eq:new_opt}, i.e.,
\begin{align}\label{eq:energy_sorting_opt}
      B_m=  \underset{B \in \mathbb{P}}{\text{argmin}} \min_{\theta \in [-\pi, \pi)}\braket{\Psi^{(m-1)}|\exp(\imath \theta \widetilde{B})H\exp(-\imath \theta \widetilde{B})|\Psi^{(m-1)}}.
\end{align}

\item Exit the iterative process if the stopping criterion is met. Otherwise, append the resulting parametrised unitary operator to the left of the current ansatz wave-function $\ket{\Psi^{(m-1)}}$, i.e., define the new ansatz wave-function
\begin{align*}
    \ket{\Psi^{(m)}}:=& \exp(-\imath \theta_m'B_m)\ket{\Psi^{(m-1)}}=\exp(-\imath \theta_m'B_m)\exp(-\imath \theta_{m-1}'B_{m-1})\ldots \exp(-\imath \theta_1'B_1)\ket{\Psi^{(0)}},
\end{align*}
where the angle $\theta_m'$ is obtained in the process of solving the optimisation problem \eqref{eq:energy_sorting_opt}.

    \item Return to Step 2 with the updated ansatz $\ket{\Psi^{(m)}}$. 

\end{enumerate}

It is important to emphasise that, in contrast to the classical ADAPT-VQE procedure, the GGA-VQE algorithm described above \emph{does not involve} a global optimisation of all parameters in the current ansatz at each iteration. Instead, at each iteration, we use the energy sorting algorithm to identify the locally optimal Hermitian generator $B$ as well as the optimal angle $\theta$ which should be used to construct the new ansatz wave-function-- a process which involves the optimisation of one-dimensional, elementary trigonometric functions. In particular, in contrast to the classical ADAPT-VQE, we avoid entirely the need to optimise a multi-dimensional and extremely noisy cost function involving the system Hamiltonian. The resulting huge savings in quantum resources suggest that the GGA-VQE is particularly suited for implementation on near-term quantum devices. Let us also note that the $d$-dimensional landscape functions introduced in Section \ref{sec:d-dimen} above can be used to develop natural generalisations of the GGA-VQE algorithm, which we dub GGA-VQE(d), that are likely to be particularly suited for the ground state preparation of strongly correlated systems.

The main computational bottle-neck in the GGA-VQE algorithm is the energy sorting procedure which requires $\mathcal{O}(M)$ evaluations of the Hamiltonian expectation energy for an operator pool of size $M$. It is therefore natural to ask if the number of measurements required to perform the energy sorting can be further reduced, at least for certain types of Hamiltonians. In the next section, we answer this question affirmatively by showing that for a certain class of Ising Hamiltonians, it is possible to perform energy sorting using a number of measurements \emph{independent of both the size of the operator pool and the number of qubits.}

\subsection{GGA-VQE for a Transverse-field Ising Model}\label{sec:Energy_Sorted_Ising}

While the ADAPT-VQE algorithm is predominantly applied to compute the ground state energies of molecular systems, there is, in principle, no restriction in applying the method to obtain ground state energies for more general Hamiltonians \cite{romero2022solving}. The goal of this section is to describe in detail, the application of the GGA-VQE algorithm that we have introduced in Section \ref{sec:Energy_Sorted} to an open boundary transverse-field Ising Hamiltonian \cite{pfeuty1970one}. Ising Hamiltonians of this type are of great importance in condensed-matter physics since they are among the simplest models capable of representing different phases of matter, depending on the value of various systems parameters \cite{sabarreto1985thermodynamical}. As the Ising Hamiltonian is well-known theoretically, it also presents a good first test for computational experiments prior to tackling more complex molecular Hamiltonians.

Given an $N$-qubit register, we consider the transverse-field Ising Hamiltonian given by
\begin{align}\label{eq:Ising_Hamiltonian}
H=h\sum_{p=0}^{N-1}X_p+J\sum_{p=0}^{N-2}Z_p Z_{p+1},
\end{align}
where $X_p$ and $Z_p$ denote the usual $X$ and $Z$ Pauli matrices acting on qubit $p$, and $h, J >0$ are system parameters. The physical constant $h$ models the intensity of a magnetic field directed along the $x$-axis, whereas the constant $J$ models the strength of the nearest-neighbour interactions. If~$J<0$, neighbouring spins tend to align, and the opposite is true if $J> 0$. Note that in this model, each qubit represents a spin-state.

Since the Ising Hamiltonian is real valued, a natural choice of operator pool is the minimal hardware-efficient pool introduced in Section \ref{sec:Pools}, which is given by
\begin{align}\label{eq:Ising_Pool}
    \mathbb P=\{Y_p\}_{p=0}^{N-1} ~\cup ~\{Z_pY_{p+1}\}_{p=0}^{N-1}.
\end{align}

Let us now recall from Section \ref{sec:Energy_Sorted} that implementing the GGA-VQE algorithm requires us to solve, at each iteration, a minimisation problem so as to identify the optimal Hermitian generator which should be used to construct the new ansatz wave-function. The objective function associated with this minimisation problem (see Equation \eqref{eq:new_opt}) is given by
\begin{align*}
  \mathcal{L}(B, \theta, \ket{\Psi^{(m-1)}})=\braket{\Psi^{(m-1})|\exp(\imath \theta B)H\exp(-\imath \theta B)|\Psi^{(m-1)}},
\end{align*}
where $B \in \mathbb{P}$ is any Hermitian generator from our operator pool, the parameter $\theta \in [-\pi,\pi)$, and $\ket{\Psi^{(m-1)}}$ denotes the previous ansatz wave-function.

It can now be shown (see the Appendix for a detailed demonstration) that for the Ising Hamiltonian defined through Equation \eqref{eq:Ising_Hamiltonian} and the minimal hardware-efficient pool $\mathbb{P}$ given by \eqref{eq:Ising_Pool}, the objective function $\mathcal{L}(B, \theta, \ket{\Psi^{(m-1)}})$ has the following simple structure:

\begin{equation}
    \mathcal{L}(B, \theta, \ket{\Psi^{(m-1)}}) = \begin{cases}
 \Big\langle \Psi^{(m-1)} \Big\vert H \Big\vert \Psi^{(m-1)}\Big \rangle \hspace{7cm}\text{ if }~ B= Y_p,\\[0.5em]
 +\sin(2\theta)\Big\langle
\Psi^{(m-1)}\Big\vert hZ_p-J(X_pZ_{p+1}+Z_{p-1}X_p\delta_{p>0})\Big\vert \Psi^{(m-1)}\Big\rangle\\[0.5em] 
-2\sin^2(\theta)\Big\langle\Psi^{(m-1)}\Big\vert hX_p+J(Z_pZ_{p+1}+Z_{p-1}Z_p\delta_{p>0})\Big\vert \Psi^{(m-1)}\Big\rangle.\\[2em]
\Big\langle \Psi^{(m-1)}  \Big\vert H \Big\vert \Psi^{(m-1)} \Big\rangle  \hspace{7cm}\text{ if }~ B= Z_pY_{p+1}. \\[0.5em]
 +\sin(2\theta)\Big\langle \Psi^{(m-1)}\Big \vert h(Z_pZ_{p+1}-Y_pY_{p+1})\Big\vert\Psi^{(m-1)}\Big \rangle\\[0.5em]
 -\sin(2\theta)\Big\langle \Psi^{(m-1)} \Big\vert J(X_{p+1}+Z_pX_{p+1}Z_{p+2}\delta_{p+2<n})\Big\vert \Psi^{(m-1)}  \Big\rangle\\[0.5em]
 -2\sin^2(\theta)\Big\langle \Psi^{(m-1)} \Big\vert hX_p+hX_{p+1}+JZ_pZ_{p+1}+JZ_{p+1}Z_{p+2}\delta_{p+2<n}\Big\vert \Psi^{(m-1)} 
 \Big\rangle.
    \end{cases}
    \label{eq:ising_landscape}
\end{equation}

\vspace{3mm}

A close study of the right-hand side of Equation \eqref{eq:ising_landscape} now indicates that many terms involving the expectation values of the Pauli matrices can be measured directly and simultaneously on the quantum device without the need to run over all possible Hermitian generators in $\mathbb{P}$.

\begin{itemize}
    \item The terms containing only tensor products of $Z$ operators can readily be measured in the computational basis.

    \item The terms containing only tensor products of $X$ (resp. $Y$) operators can be measured by applying a Hadamard (resp. $S^\dag\equiv \text{diag}(1,-\imath)$ and a Hadamard) gate on each qubit.

    \item  The remaining terms are of the form $X_pZ_{p+1}$ or $Z_{p-1}X_{p}Z_{p+1}$. Terms of this form can be measured by applying a Hadamard gate on qubit $p$. The terms corresponding to $p$ even commute and can therefore be measured simultaneously. The same holds true for the $p$ odd terms which can thus also be measured simultaneously.
\end{itemize}
   
Consequently, it is possible, at each iteration of the Greedy-ADAPT-VQE algorithm to construct exactly five quantum circuits whose measurements allow us to recover an analytical expression for all objective functions $\mathcal{L}(B, \theta, \ket{\Psi^{(m-1)}}), ~B\in \mathbb{P}$ in terms of elementary trigonometric functions of $\theta$. We have thus achieved a radical reduction in the number of required measurements from $4N-3$ (for the minimal hardware-efficient pool of size $2N-2$) to \emph{five}. \vspace{3mm}

We end this section by noting that a simple choice of initial state for the energy-sorting frozen core ADAPT-VQE procedure is given by the ground-state of the non-interacting Hamiltonian $\sum_{p=0}^{N-1} X_p$, i.e.,
\begin{align*}
\ket{\Psi^{(0)}}=\ket-^{\otimes n}
\end{align*}
Assuming that the system parameters satisfy $|h|>|J|$, it is not unreasonable to expect the ground state of the true \emph{interacting} Hamiltonian to be a perturbation of $\ket{\Psi^{(0)}}$.

\subsection{Hybrid Measurements of Quantum Simulations}

The objective of adaptive algorithms such as GGA-VQE or ADAPT-VQE is to obtain an ordered set of parameterized operators  whose application to the initial state yields an accurate approximation of the ground state. Typically, the accuracy of the approximation is evaluated by comparing the energy of the ansatz wave-function to the true ground state energy. 
When executing such algorithms on quantum architectures, if we observe a large deviation between the ansatz wave-function energy and the true ground state energy, we can classify the reasons for this discrepancy as follows:
\begin{enumerate}
    \item[(\textbf{R1})] The device and measurement (statistical) noise on the quantum device results in sub-optimal choices of parameterised operators and corresponding parameter values;
    \item[(\textbf{R2})] Even if the choice of parameterised operators and corresponding parameter values are satisfactory, the device and measurement noise on the quantum device prevents an accurate evaluation of the ansatz wave-function energy;
    \item[(\textbf{R3})] Even if the choice of parameterised operators and corresponding parameter values are satisfactory and the ansatz wave-function energy can be accurately evaluated, the adaptive algorithm itself does a poor job at approximating the ground state of the system under study.
\end{enumerate}

Typically, (\textbf{R3}) above is studied by executing the adaptive algorithms in noise-free, idealised settings, and such studies have led to improved choices of operator pools \cite{tang2021qubit} as well as more sophisticated adaptive algorithms \cite{burton2023exact}. These developments are beyond the scope of the present work however. Indeed, since our motivation is the effect of measurement and device noise on the success of adaptive quantum algorithms, we will, as a first step, focus instead on (\textbf{R1}) above. More precisely, we will evaluate whether the QPU implementations of our adaptive algorithms yield satisfactory (although perhaps not optimal) choices of parameterised operators and corresponding parameter values used to define the ansatz wave-function.  

An obvious strategy to make this evaluation, is to retrieve the parameterised operators and corresponding parameter values yielded by the QPU-executed GGA-VQE method, represent the corresponding ansatz wave-function on a noiseless HPC emulator, and measure the sought-after observables. We refer to this approach as 'hybrid' observable evaluation in the sequel. It is important to emphasise that the construction of the ansatz wave-function, i.e., the choice and order of unitary operators as well as the corresponding angles \underline{are all determined by calculations on the QPU}. It is only the final measurement of the ansatz wave-function that takes place on the emulator. This hybrid observable measurement is, admittedly, not a scalable procedure (since it requires the emulation of quantum circuits on classical HPC architectures). However, it is our opinion that in the present era of NISQ devices, it can be used as a tool to distinguish between (\textbf{R1}) and (\textbf{R2})-type failures of adaptive quantum algorithms on QPUs.

\begin{figure}[H]
    \centering
    \includegraphics[width=0.7\textwidth]{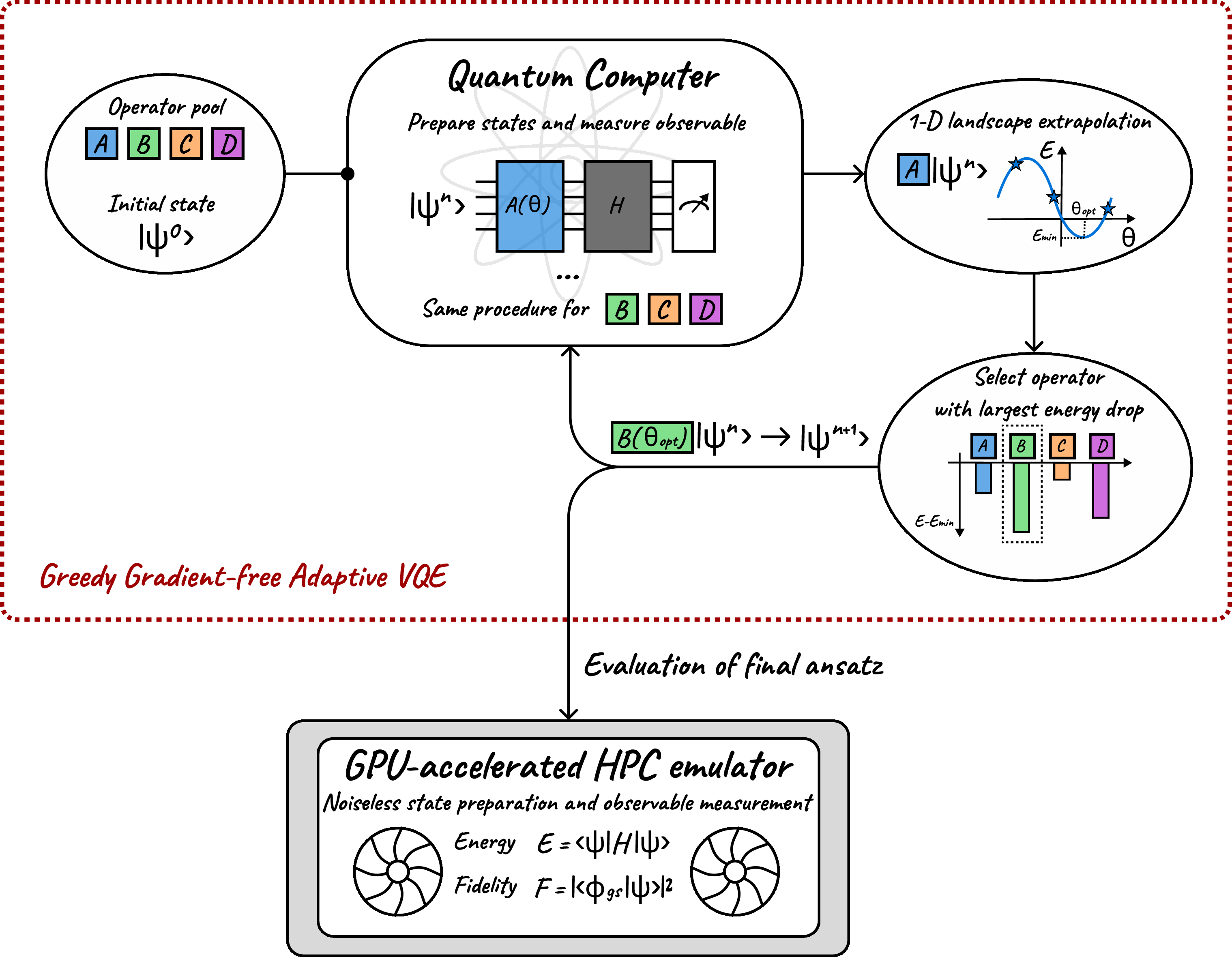}
    \caption{A schematic overview of the GGA-VQE algorithm with hybrid observable measurement. We emphasise that the HPC emulator is used \emph{only} to evaluate the final ansatz wave-function obtained by executing GGA-VQE on the QPU.}
\end{figure}


\section*{Data availability}
Data generated during the study is available upon request from the authors (E-mail:
jean-philip.piquemal@sorbonne-universite.fr).
\section*{Code availability}
The code used during the study is available upon request from the authors (E-mail:
jean-philip.piquemal@sorbonne-universite.fr).
\section*{Competing interests}
JPP is shareholder and co-founder of Qubit Pharmaceuticals. The remaining authors declare no other competing interests.
\section*{Acknowledgements}

The authors wish to thank three anonymous referees for their suggestions and comments. This work has been funded by the European Research Council (ERC) under the European Union’s Horizon 2020 research and innovation program (grant No 810367), project EMC2 (J.-P. P. and Y.M.). Support from the PEPR EPIQ - Quantum Software (ANR-22-PETQ-0007, JPP) and HQI (Y.M, JPP) programs is acknowledged. We thank Amazon Braket (G. Tourpe) for partial funding of the computations on the Aria IonQ machine and the whole team for their help in the setup of our computations on Amazon Braket's SDK (Software Development Kit). GPU computations have been performed at GENCI (IDRIS, Orsay, France) on grant no A0130712052.

\section{Appendix}

\subsection{Quantum circuits for qubit-excitation operators}~
\label{qeb_circuits}
\begin{figure}[h]
\centering
\includegraphics[scale=0.35]{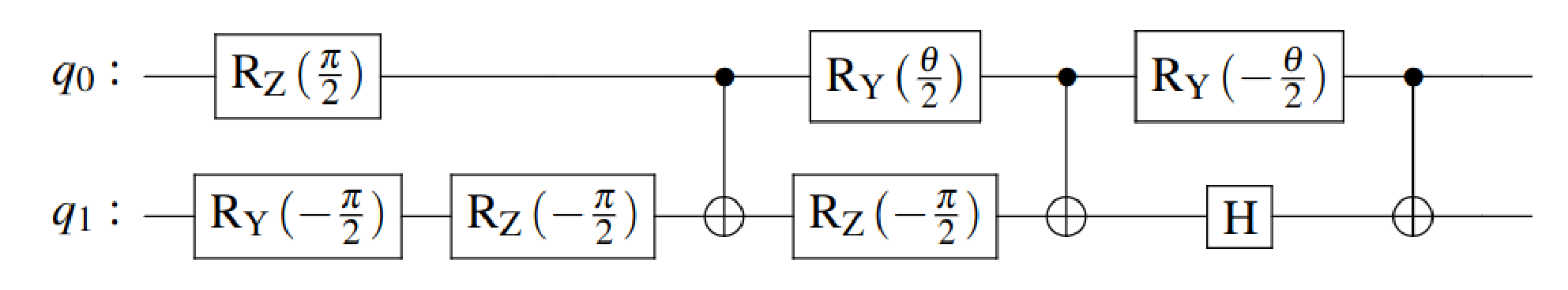}
\caption{A quantum circuit performing a generic single-qubit evolution \cite{Yordanov_2020}.}
\label{qcirc:sqe}
\end{figure}

\begin{figure}[ht]
\centering
\includegraphics[scale=0.3]{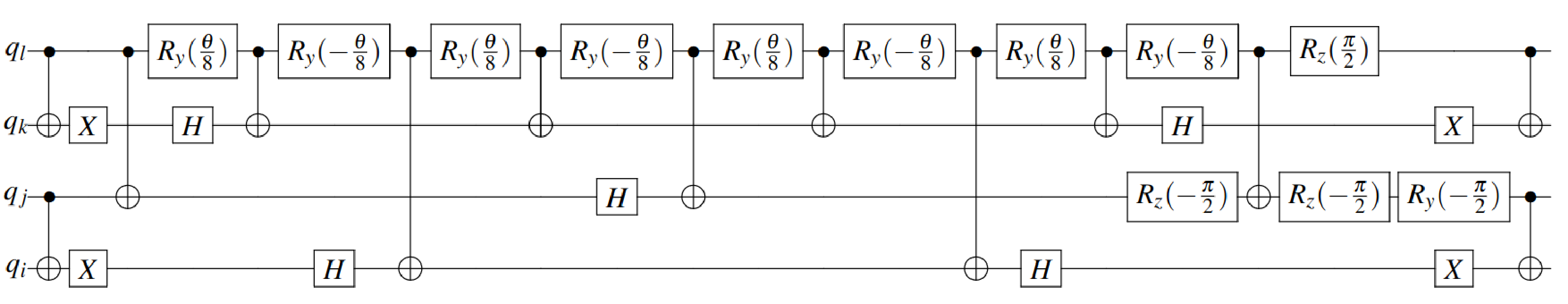}
\caption{A quantum circuit performing a generic double-qubit evolution \cite{Yordanov_2020}.}
\label{fig:d_q_exc_full}
\end{figure}

For the sake of completeness, we present a few key quantum circuits used in the hardware experiments carried out for this study. The circuit for a single-qubit excitation is given in Figure \ref{qcirc:sqe} whereas the circuit for a double-qubit excitation is displayed in Figure \ref{fig:d_q_exc_full}. Both qubit excitations correspond to the qubit-excitation based (QEB) pool introduced in Section \ref{sec:Pools}, and, as explained in \cite{Yordanov_2020}. Note that although we used these quantum circuits for the QEB operators, we have since been made aware that more hardware-efficient circuits have been developed \cite{yordanov2021qubit, sun2024circuit}.

Remark: when displaying quantum circuits, $H$ denotes the Hadamard gate and not a physical Hamiltonian.

\subsection{Periodicity of QEB operator pool and involutory property of hardware-efficient pools}\label{appendix_pools}~

Recall the definition of the qubit excitation-based (QEB) pool given in Section \ref{sec:Pools} and $A_{pqrs}$ denote a double-qubit excitation generator between the pairs of qubits $(p,q)$ and $(r,s)$ as defined through Equation \eqref{eq:double}, i.e.,
\begin{equation*}
\begin{split}
 A_{pqrs}=\frac{1}{8}&\left(
X_rY_sX_pX_q+Y_rX_sX_pX_q+Y_rY_sY_pX_q+Y_rY_sX_pY_q -X_rX_sY_pX_q-X_rX_sX_pY_q-Y_rX_sY_pY_q-X_rY_sY_pY_q\right). 
\end{split}
\end{equation*}

A direct calculation shows that $A_{pqrs}$ can be written in the equivalent form
\begin{align*}
A_{pqrs}=i(Q_p^\dag Q_q^\dag Q_rQ_s-Q_r^\dag Q_s^\dag Q_pQ_q),
\end{align*}
where for any qubit index $a$, we define $Q_a=\frac12(X_a+iY_a)$. 

Using this representation, we can easily show that $A_{pqrs}\ket{1_p1_q0_r0_s}=i\ket{0_p0_q1_r1_s}$, $A_{pqrs}\ket{0_p0_q1_r1_s}=-i\ket{1_p1_q0_r0_s}$ while the action of $A_{qprs}$ on all other states is zero. Consequently, the subspace spanned by $e_1\equiv\ket{1_p1_q0_r0_s}$ and $e_2\equiv\ket{0_p0_q1_r1_s}$ is an invariant subspace of $A_{pqrs}$, and in the basis $\{e_1,e_2\}$ of this invariant subspace, $A_{pqrs}$ has the representation
\begin{align*}
A_{pqrs}=\begin{pmatrix}
0&-\imath\\\imath&0
\end{pmatrix},
\end{align*}
which is the well-known $Y$ Pauli matrix. We thus conclude that $A_{pqrs}$ has eigenvalues $0,\pm1$ and satisfies $A_{pqrs}^3=A_{pqrs}$. 

A similar demonstration can be carried out for single-qubit generators from the QEB pool and generators from the qubit hardware efficient and minimal hardware efficient pools, which shows that these generators are involutory, i.e., they satisfy $B^2=I$. For the sake of brevity, we do not provide a detailed argument.

The above observation motivates further investigation and leads to the following result.

\begin{theorem}
Let $H$ denote an N-qubit Hamiltonian and let $\mathbb{P}$ denote any of the operator pools introduced in Section \ref{sec:Pools}. Then define, for any $N$-qubit wave-function $\ket{\phi}$, any Hermitian generator $B \in \mathbb{P}$ and any $\theta \in [-\pi, \pi)$ the landscape function
\begin{align*}
\mathcal{L}\left(B, \theta, \ket{\phi} \right)= \left \langle \phi \left\vert \exp(\imath \theta B)H \exp(-\imath \theta B)\right\vert \phi \right \rangle.    
\end{align*}
Then it holds that
 \begin{equation*}
\mathcal{L}(B, \theta, \ket{\phi}) =\begin{cases} \braket{\phi|H|\phi}
+\big(\cos(\theta)-1\big)(\braket{\phi|\{H,B^2\}|\phi}-2\braket{\phi|BHB|\phi} ) \hspace{2cm} \text{if } B^3=B,\\[0.5em]
+\big(1-\cos(\theta)\big)^2\big(\braket{\phi|B^2HB^2|\phi}-\braket{\phi|BHB|\phi} \big)\\[0.5em]
+\sin(\theta)(\cos(\theta)-1)\braket{\phi|\imath B[H,B]B|\phi} \\[0.5em]
+\sin(\theta)\braket{\phi|\imath[B,H]|\phi}\\[2em]
\cos^2(\theta)\braket{\phi|H|\phi} +\dfrac{\sin(2\theta)}{2}\braket{\phi|\imath [B,H]|\phi} \hspace{4.15cm}\text{if } B^2=I.\\[0.75em]
+\sin^2(\theta)\braket{\phi|BHB|\phi},
\end{cases}
\end{equation*}
where $\{\cdot, \cdot\}$ and $[\cdot, \cdot]$ denote the anti-commutator and commutator respectively.
\end{theorem}
\begin{proof}
We consider first the case $B^3=B$. For such a Hermitian generator, we can use the Taylor series expansion of the exponential to deduce that
\begin{equation}\label{eq:appendix_new_1}
\begin{split}
\exp(-\imath\theta B)&=\sum_{k=0}^\infty\frac{(-i\theta B)^{2k}}{(2k)!}+\sum_{k=0}^\infty\frac{(-i\theta B)^{2k+1}}{(2k+1)!}\\[0.5em]
&=I+(\cos(\theta)-1)B^2-\imath\sin(\theta) B.
\end{split}
\end{equation}
Plugging in the expression \eqref{eq:appendix_new_1} into the definition of the landscape function $\mathcal{L}\left(B, \theta, \ket{\phi} \right)$ now yields the desired result. The case $B^2=I$ is simply a special case.
\end{proof}

\subsection{Analytical expressions of GGA-VQE objective functions for the Ising Hamiltonian}~

Throughout this section, we use the setting and notation of Section \ref{sec:Energy_Sorted_Ising}. Our goal now is to demonstrate that for the Ising Hamiltonian defined through Equation \eqref{eq:Ising_Hamiltonian} and the minimal hardware-efficient pool $\mathbb{P}$ given by \eqref{eq:Ising_Pool}, the objective function $\mathcal{L}(B, \theta, \ket{\Psi^{(m-1)}})$ has the following simple structure:
\begin{equation}\label{eq:appendix_Ising}
    \mathcal{L}(B, \theta, \ket{\Psi^{(m-1)}}) = \begin{cases}
 \Big\langle \Psi^{(m-1)} \Big\vert H \Big\vert \Psi^{(m-1)}\Big \rangle \hspace{7cm}\text{ if }~ B= Y_p,\\[0.5em]
 +\sin(2\theta)\Big\langle
\Psi^{(m-1)}\Big\vert hZ_p-J(X_pZ_{p+1}+Z_{p-1}X_p\delta_{p>0})\Big\vert \Psi^{(m-1)}\Big\rangle\\[0.5em] 
-2\sin^2(\theta)\Big\langle\Psi^{(m-1)}\Big\vert hX_p+J(Z_pZ_{p+1}+Z_{p-1}Z_p\delta_{p>0})\Big\vert \Psi^{(m-1)}\Big\rangle.\\[2em]
\Big\langle \Psi^{(m-1)}  \Big\vert H \Big\vert \Psi^{(m-1)} \Big\rangle  \hspace{7cm}\text{ if }~ B= Z_pY_{p+1}. \\[0.5em]
 +\sin(2\theta)\Big\langle \Psi^{(m-1)}\Big \vert h(Z_pZ_{p+1}-Y_pY_{p+1})\Big\vert\Psi^{(m-1)}\Big \rangle\\[0.5em]
 -\sin(2\theta)\Big\langle \Psi^{(m-1)} \Big\vert J(X_{p+1}+Z_pX_{p+1}Z_{p+2}\delta_{p+2<n})\Big\vert \Psi^{(m-1)}  \Big\rangle\\[0.5em]
 -2\sin^2(\theta)\Big\langle \Psi^{(m-1)} \Big\vert hX_p+hX_{p+1}+JZ_pZ_{p+1}+JZ_{p+1}Z_{p+2}\delta_{p+2<n}\Big\vert \Psi^{(m-1)} 
 \Big\rangle.
    \end{cases}
\end{equation}

To show that Equation \eqref{eq:appendix_Ising} indeed holds, we first recall the definition of the objective function $ \mathcal{L}(B, \theta, \ket{\Psi^{(m-1)}})$ which is given by
\begin{align}\label{eq:appendix_ising_1}
  \mathcal{L}(B, \theta, \ket{\Psi^{(m-1)}})=\braket{\Psi^{(m-1})|\exp(\imath \theta B)H\exp(-\imath \theta B)|\Psi^{(m-1)}},
\end{align}
where $B \in \mathbb{P}$ is any Hermitian generator from the minimal hardware-efficient operator pool, the parameter $\theta \in [-\pi,\pi)$, and $\ket{\Psi^{(m-1)}}$ denotes the previous ansatz wave-function. 

Next, we recall from Equation \eqref{eq:landscape} that the involutory property of the Hermitian generators from the minimal hardware-efficient pool yields the following simplification of Equation \eqref{eq:appendix_ising_1}:
\begin{align}\label{eq:appendix_ising_2}
  \mathcal{L}(B, \theta, \ket{\Psi^{(m-1)}})=\cos^2(\theta)\braket{\phi|H|\phi} +\dfrac{\sin(2\theta)}{2}\braket{\phi|\imath [B,H]|\phi} +\sin^2(\theta)\braket{\phi|BHB|\phi}.
\end{align}
Consequently, in order to arrive at Equation \eqref{eq:appendix_Ising}, we have to simplify each term involving the Ising Hamiltonian and minimal hardware-efficient generator $B$ appearing in Equation \eqref{eq:appendix_ising_2}.

To do so, recall that we denote the total number of qubits (i.e., the size of the quantum register) by $N \in \mathbb{N}$, and fix an index $p \in \{0, \ldots, N-2\}$. Using now the commutation relations of the Pauli matrices, a direct calculation reveals that
 \begin{align}\nonumber
 [Y_p,H]&=-2h\imath Z_p+2\imath J(X_p Z_{p+1}+Z_{p-1}X_p\delta_{p>0})\\
 \intertext{and} 
 [Z_p Y_{p+1},H]&=2\imath h(Y_p Y_{p+1}-Z_p Z_{p+1})+2\imath J(X_{p+1}+Z_p X_{p +1}Z_{p+2}\delta_{p<N-2}). \label{eq:appendix_ising_3}
 \end{align}
 
 A similar calculation utilising once again the commutation relations of the Pauli matrices further yields that
 \begin{align}\nonumber
 Y_pHY_p&= H-2h X_p-2 J (Z_pZ_{p+1}+Z_{p-1}Z_j\delta_{p>0}),
 \intertext{and} 
 Z Z_{p+1} H Z_p Z_{p+1}&=\sum_{q=0}^{N-1} Z_p Z_{p+1} X_q Z_p Z_{p+1} = H-2h (X_p+X_{p+1}). \label{eq:appendix_ising_4}
 \end{align}

The result now follows by plugging in Equations \eqref{eq:appendix_ising_3} and \eqref{eq:appendix_ising_4} into Equation \eqref{eq:appendix_ising_2}.
 

 


\vspace{5mm}
\subsection{Reducing the computational complexity of the energy sorting algorithm for general spin chains}\label{sec:general_spin}~

As demonstrated in Section \ref{sec:Energy_Sorted_Ising}, the specific structure of transverse-field Ising Hamiltonian leads to a huge reduction in the computational cost of the energy sorting step of the GGA-VQE algorithm. Indeed, while the energy sorting step a priori requires $\mathcal{O}(M)$ measurements for a general system Hamiltonian and an operator pool of size $M$, the number of required measurements reduces to just \emph{five} in the case of the one-dimensional transverse field Hamiltonian. The goal of this section is to briefly describe similar reductions in the computational complexity of the energy sorting algorithm for Ising spin-chain Hamiltonians with local magnetic fields and couplings in all three spatial directions, i.e., Hamiltonians of the form
\begin{align}\label{eq:Ising_general}
    H = \sum_{k=0}^{N-1} h^{x}_k X_k+\sum_{k=0}^{N-1} h^{z}_k Z_k+\sum_{k=0}^{N-2} J^{x}_kX_kX_{k+1}+\sum_{k=0}^{N-2} J^{y}_k Y_kY_{k+1}+\sum_{k=0}^{N-2} J_k^zZ_kZ_{k+1}.
\end{align}
Here, $h^{x}_k$ and $h^{z}_k$ denote constants that model the intensity of the magnetic field along the $x$ and $z$ directions while $J_k^x,J_k^y$ and $J_k^z$ are constants that model the strength of the nearest-neighbour interactions in the $x, y, $ and $z$ directions respectively. \vspace{2mm}

Tables \ref{magnetic_fields} and \ref{couplings} list the terms of interest that appear in the one-dimensional GGA-VQE landscape functions that are used to perform the energy sorting step. Comparing the terms that appear in Tables \ref{magnetic_fields} and \ref{couplings} with the simpler expressions for the transverse-field Ising Hamiltonian from Section \ref{sec:Energy_Sorted_Ising}, we see that the only new terms that arise are of the $Z_{p-2}Z_{p-1}X_p$ and $Y_{p-1}X_p$. As before, we can simultaneously measure such operators acting on a disjoint set of qubits-- a process that will require an additional five quantum circuits at each step. Consequently, applying the GGA-VQE algorithm to general Ising Hamiltonians of the form \eqref{eq:Ising_general} will require constructing and measuring at most ten quantum circuits, irrespective of the number of qubits and the size of the minimal operator pool. \vspace{2mm}

\begin{table}[htp]
\centering
\begin{center}
\begin{tabular}{|c|c|c|c|}
\hline
$\mathbf{H}$ & $\mathbf{\sum h_k X_k}$&$\mathbf{\sum h_k Y_k}$&$\mathbf{\sum h_k Z_k}$\\
\hline
$\mathbf{[Y_i, H]}$ &$-2ih_iZ_i$&$0$&$2ih_iX_i$\\
\hline
$\mathbf{Y_iHY_i}$&$H-2h_iX_i$&$H$&$H-2Z_i$\\
\hline
$\mathbf{[Z_iY_{i+1},H]}$&$-2ih_iZ_iZ_{i+1}+2ih_{i+1}Y_iY_{i+1}$&$-2ih_iX_iY_{i+1}$&$2ih_{i+1}Z_iX_{i+1}$\\
\hline
$\mathbf{Z_iY_{i+1}HZ_iY_{i+1}}$ & $H-2h_{i+1}X_{k+1}-2h_{i}X_i$&$H-2h_{i+1}Y_{i+1}$&$H-2h_iZ_i-2h_{i+1}Z_{i+1}$\\
\hline
\end{tabular}
\end{center}
\caption{Commutators involving generators from the minimal operator pool and the local magnetic field terms.}
\label{magnetic_fields}
\end{table}%
 
 \begin{table}[htp]
 \begin{center}
\begin{tabular}{|c|c|c|c|}
\hline
$\mathbf{H}$ &$\mathbf{\sum J_k X_kX_{k+1}}$&$\mathbf{\sum J_k Y_kY_{k+1}}$&$\mathbf{\sum J_k Z_kZ_{k+1}}$\\
\hline
$\mathbf{[Y_i, H]}$ &$-2iJ_iZ_iX_{i+1}-2iJ_{i+1}X_iZ_{i+1}$&$0$&$2iJ_iX_iZ_{i+1}+2iJ_{i-1}Z_{i-1}X_i$\\
\hline
$\mathbf{Y_iHY_i}$&$H-2J_iX_iX_{i+1}-2J_{i+1}X_{i-1}X_i$&$H$&$H-2J_iZ_iZ_{i+1}-2J_{i-1}Z_{i-1}Z_i$\\
\hline
$\mathbf{[Z_iY_{i+1},H]}$&$-2iJ_{i+1}Z_iZ_{i+1}X_{i+2}-2iJ_{i-1}X_{i-1}Y_{i}Y_{i+1}$&$-2iJ_iX_i-2iJ_{i-1}Y_{i-1}X_iY_{i+1}$&$2iJ_{i}X_{i+1}+2iJ_{i+1}Z_iX_{i+1}Z_{i+2}$\\
\hline
$\mathbf{Z_iY_{i+1}HZ_iY_{i+1}}$ & $H-2J_{i-1}X_{i-1}X_i$&$H-2J_iY_{i}Y_{i+1}-2J_{i-1}Y_{i-1}Y_{i}$&$H-2J_{i+1}Z_{i+1}Z_{i+2}-2J_iZ_iZ_{i+1}$\\
\hline
\end{tabular}
\caption{Commutators involving generators from the minimal operator pool and the interaction terms in each direction.}
\label{couplings}
\end{center}
\end{table}

Finally, let us remark that we expect similar but likely less drastic simplifications to also hold for Hamiltonians arising from other physical models.

\vspace{5mm}

\subsection{Multi-dimensional analytical landscape functions and post-processing of GGA-VQE}\label{sec:d-dimen}~\vspace{2mm}

Let us consider an adaptive procedure in which $d$ unitary operators, constructed using $d$ Hermitian generators from a given operator pool $\mathbb{P}$ are to be appended to the current ansatz wave-function $\ket{\Psi^{(m-1)}}$ at iteration $m$. We are now interested in determining the ordered $d$-tuple of Hermitian generators $(B_{m_d}, \ldots, B_{m_1})$ such that
\begin{align}\label{eq:new_opt_general}
    (B_{m_d}, \ldots, B_{m_1})=& \underset{B_d, \ldots, B_1 \in \mathbb{P}} {\text{argmin}}  \hspace{2mm}\min_{\substack{\theta_d, \ldots \theta_1 \\[0.2em] \in [-\pi, \pi)}}\mathcal{L}\Big( (B_d, \theta_d),\ldots, (B_1, \theta_1), \ket{\Psi^{(m-1)}}\Big)\\ \nonumber
   :=&  \underset{B_d, \ldots, B_1 \in \mathbb{P}}{\text{argmin}} \hspace{2mm} \min_{\substack{\theta_d, \ldots \theta_1\\[0.2em] \in [-\pi, \pi)}} \braket{\Psi^{(m-1)}|\exp(\imath \theta_1 {B}_1)\ldots \exp(\imath \theta_d B_d) H\exp(-\imath \theta_d B_d)\ldots\exp(-\imath \theta_1 B_1)|\Psi^{(m-1)}}.
\end{align}

In order to obtain an analytical representation of the $d$-dimensional objective function $\mathcal{L}\Big( (B_d, \theta_d),\ldots, (B_1, \theta_1), \ket{\Psi^{(m-1)}}\Big)$, the fundamental idea is to appeal once again to Equations \eqref{eq:new_has_1} and \eqref{eq:new_has_2} and expand each exponential in $\theta_j, ~ j \in \{1, \ldots, d\}$ as a sum of a sine and cosine function of $\theta_j$. This expansion allows us to conclude that the  $d$-dimensional objective function $\mathcal{L}\Big( (B_d, \theta_d),\ldots, (B_1, \theta_1), \ket{\Psi^{(m-1)}}\Big)$ can be written in the general form
\begin{align*}\label{eq:landscape_2}
&\mathcal{L}\Big( (B_d, \theta_d),\ldots, (B_1, \theta_1), \ket{\Psi^{(m-1)}}\Big)= \\ \nonumber
&\begin{cases}  \Bigg \langle \Psi^{(m-1)}\Big \vert \underset{j=1}{\overset{j=d}{\prod}} \left(I+(\cos(\theta_j)-1)B_j^2+\imath\sin(\theta_j) B_j\right) H\underset{j=d}{\overset{j=1}{\prod}} \left(I+(\cos(\theta_j)-1)B_j^2-\imath\sin(\theta_j) B_j\right)\Big \vert \Psi^{(m-1)}\Bigg \rangle \hspace{0.7cm} \text{if } B^3=B,\\[2em]
 \Bigg \langle \Psi^{(m-1)}\Big \vert \underset{j=1}{\overset{j=d}{\prod}} \left(\cos(\theta_j)I+\imath\sin(\theta_j) B_j\right) H\underset{j=d}{\overset{j=1}{\prod}} \left(\cos(\theta_j)I-\imath\sin(\theta_j) B_j\right)\Big \vert \Psi^{(m-1)}\Bigg \rangle\hspace{4.15cm}\text{if } B^2=I.
\end{cases}
\end{align*}


In other words, the $d$-dimensional objective function $\mathcal{L}\Big( (B_d, \theta_d),\ldots, (B_1, \theta_1), \ket{\Psi^{(m-1)}}\Big)$ can be written as a polynomial of the variables $\big\{1, \cos(\theta_j), \sin(\theta_j)\colon j \in \{1, \ldots, d\} \big\}$ with the exact structure of the polynomial depending on the properties of the operator pool $\mathbb{P}$. As a representative example, the landscape function for hardware efficient pools with $d=2$, after some simplifications, is of the form:

\begin{equation}
\label{eq:landscape_2}
\begin{split}
\mathcal{L}\big((B_2, \theta_2), (B_1, \theta_1), \ket{\phi}\big) =&
 \braket{\phi|H|\phi}\\[0.25em]
 +&\frac{\cos(2\theta_1)}{2}\Big(\braket{\phi| H-B_1HB_1|\phi} +\frac{\cos(2\theta_2)}2\braket{\phi| H-B_1HB_1-B_2HB_2+B_2B_1HB_1B_2|\phi}\\[0.25em]
&\hspace{1cm}+\frac{\sin(2\theta_2)}2\braket{\phi| i[B_2,H-B_1HB_1]|\phi}\Big)\\
+&\frac{\sin(2\theta_1)}2\Big(
\langle\phi| i[B_1,H]|\phi\rangle+\frac{\cos(2\theta_2)}2\langle \phi|
i[B_1,H]-iB_2[B_1,H]B_2|\phi\rangle\\[0.25em]
&\hspace{1cm}-\frac{\sin(2\theta_2)}2\langle\phi|[B_2,[B_1,H]]|\phi\rangle\Big)
\end{split}
\end{equation}

Consequently, a total of 7 observable evaluations on a quantum device are required to deduce an analytical expression for the two-dimensional landscape function $\mathcal{L}\big((B_2, \theta_2), (B_1, \theta_1), \ket{\phi}\big)$ for any Hermitian generators $B_1, B_2$ belonging to either of the two hardware efficient operator pools. Since the selection of the best two operator to append to the current ansatz wave-function requires comparing all pairs of Hermitian generators, we conclude that for an operator pool of size $M$, at most $6M^2+1$ measurements are required to determine the locally optimal pair of unitary operators that should be appended to the current ansatz wave-function at each iteration in order to achieve the largest drop in expectation value of the underlying Hermitian operator.

In the case of a general $d$-dimensional objective function $\mathcal{L}\Big( (B_d, \theta_d),\ldots, (B_1, \theta_1), \ket{\Psi^{(m-1)}}\Big)$, similar arguments yield that
\begin{itemize}
    \item for the Qubit-Excitation-Based (QEB) operator pool of size $M$, we require $\mathcal{O}(5^d M^d)$ measurements to determine the locally optimal $d$-tuple of unitary operators that should be appended to the current ansatz wave-function at each iteration in order to achieve the largest drop in expectation value of the underlying Hermitian operator;

    \item for the Qubit hardware-efficient and minimal hardware-efficient pools of size $M$, we require $\mathcal{O}(3^d M^d)$ measurements to determine the locally optimal $d$-tuple of unitary operators that should be appended to the current ansatz wave-function at each iteration in order to achieve the largest drop in expectation value of the underlying Hermitian operator.    
\end{itemize}

While this procedure can become computationally intractable for moderately large $d$, various simplifications are possible that can lead to more tractable gradient-free adaptive algorithms involving multi-operator selection and optimisation, and it is likely that such methods offer an advantage when formulating a greedy gradient-free adaptive VQE for a complex Hamiltonian using a non-commutative operator pool. As representative examples, given an ansatz wave-function $\ket{\Psi^{(m-1)}}$, at iteration $m$:

\begin{enumerate}
    \item We may use the energy sorting algorithm based on \emph{one-dimensional landscape functions} to classify, in descending order of importance, the best $d$ Hermitian generators $(B_d, \ldots, B_1)$ whose addition to the current ansatz wave-function can result in the largest drops in the expectation value of the underlying Hermitian operator, i.e.,
    \begin{align*}
        \min_{\theta_d \in [-\pi, \pi)}\braket{\Psi^{(m-1)}|\exp(\imath \theta B_d)H\exp(-\imath \theta B_d)|\Psi^{(m-1)}} \geq \ldots \geq \min_{\theta_1 \in [-\pi, \pi)}\braket{\Psi^{(m-1)}|\exp(\imath \theta B_1)H\exp(-\imath \theta B_1)|\Psi^{(m-1)}}.
    \end{align*}
    
    We can then switch to the analytical expression of $d$-dimensional landscape function $\mathcal{L}\Big( (B_d, \theta_d),\ldots, (B_1, \theta_1), \ket{\Psi^{(m-1)}}\Big)$ in order to compute the optimal parameters $(\theta_d^*, \ldots, \theta^*_1)$ such that 
    \begin{align*}
    (\theta_d^*, \ldots, \theta_1^*)=&\text{argmin}_{\substack{\theta_d, \ldots \theta_1 \\[0.2em] \in [-\pi, \pi)}}\mathcal{L}\Big( (B_d, \theta_d),\ldots, (B_1, \theta_1), \ket{\Psi^{(m-1)}}\Big)\\
    =&\text{argmin}_{\substack{\theta_d, \ldots \theta_1 \\[0.2em] \in [-\pi, \pi)}}\braket{\Psi^{(m-1)}|\exp(\imath \theta_1 {B}_1)\ldots \exp(\imath \theta_d B_d) H\exp(-\imath \theta_d B_d)\ldots\exp(-\imath \theta_1 B_1)|\Psi^{(m-1)}}
    \end{align*}
    In other words, we may use one-dimensional landscape functions to \emph{identity} the best $d$ Hermitian generators to add to the current ansatz wave-function and we may employ the $d$-dimensional landscape functions to perform the \emph{analytical optimisation}. The number of quantum measurements required by this procedure scales as $3^d M$ (resp. $5^d M$) for a hardware efficient (resp. QEB) operator pool of size $M$.
    
    \item Taking the newly obtained ansatz wave-function $\ket{\Psi^{(m)}}$ after iteration $m$ as structurally fixed, we may perform Rotoselect-style \cite{ostaszewski2021structure} backwards and forwards sequential optimisation sweeps over all parameterised unitary operators $\exp(\imath \theta_1^* {B}_d), \ldots \exp(\imath \theta_d^* {B}_1)$. In particular, thanks to the analytical expression \eqref{eq:landscape_2} for the $d$-dimensional landscape function, each iteration in these optimisation sweeps can involve $d$ parametrised unitary operators simultaneously. The number of quantum measurements required by a single sweep utilising  $d$-dimensional landscape functions scales as $3^d M$ (resp. $5^d M$) for a hardware efficient (resp. QEB) operator pool of size $M$. Algorithm \ref{alg:1} below (c.f., the RotoSolve algorithm \cite{ostaszewski2021structure}) yields a simple example of such a procedure for the case $d=1$.
\end{enumerate}

    \begin{algorithm}
\caption{Sequential Reoptimisation of GGA-VQE Ansatz}\label{alg:1}
\begin{algorithmic}
\Require {At iteration $m$, we are given an ansatz $\ket{\Psi^{(m)}}=\exp(-\imath \widetilde{\theta_m} B_m)\ldots\exp(-\imath \widetilde{\theta_1} B_1)|\Psi^{(0}\rangle$.} \vspace{2mm}
\State {\textbf{Initialise} j=m-1.} \vspace{2mm}

\While{$j\geq 1$}

\State  {\textbf{(i)}~ Using measurements on the quantum device, \textbf{construct} the one-dimensional landscape function 
\begin{align*}
    \mathcal{L}_j(\theta):=\Big \langle \Psi^{(0)}\Big|\exp(\imath \widetilde{\theta_1} {B}_1)\ldots &\exp(\imath \widetilde{\theta_{j-1}} B_{j-1}) \exp(\imath \theta B_j) \exp(\imath \widehat{\theta_{j+1}} B_{j+1}) \ldots \exp(\imath \widehat{\theta_m} B_m)\\
    &H\exp(-\imath \widehat{\theta_m} B_m) \ldots  \exp(-\imath \widehat{\theta_{j+1}} B_{j+1}) \exp(-\imath \theta B_j) \exp(-\imath \widetilde{\theta_{j-1}} B_{j-1})\ldots \exp(-\imath \widetilde{\theta_1} {B}_1)\Big |\Psi^{(0}\Big \rangle
\end{align*}
}
\State {\textbf{(ii)}~ \textbf{Solve} the minimisation problem 
\begin{align*}
    \widehat{\theta_j}:= \argmin_{\theta \in [-\pi, \pi)} \mathcal{L}_j(\theta),
\end{align*}}
\State \textbf{(iii)}~ \textbf{Update} $\widetilde{\theta_j} \mapsto \widehat{\theta_j}$ and \textbf{update} $j\mapsto j-1$.\vspace{2mm}

\EndWhile\\

\While{{$j\leq m$}}

\State{\textbf{(i)}~  Using measurements on the quantum device, \textbf{construct} the one-dimensional landscape function 
\begin{align*}
    \mathcal{L}_j(\theta):=\Big \langle \Psi^{(0)}\Big|\exp(\imath \widehat{\theta_1} {B}_1)\ldots &\exp(\imath \widehat{\theta_{j-1}} B_{j-1}) \exp(\imath \theta B_j) \exp(\imath \widetilde{\theta_{j+1}} B_{j+1}) \ldots \exp(\imath \widetilde{\theta_m} B_m)\\
    &H\exp(-\imath \widetilde{\theta_m} B_m) \ldots  \exp(-\imath \widetilde{\theta_{j+1}} B_{j+1}) \exp(-\imath \theta B_j) \exp(-\imath \widehat{\theta_{j-1}} B_{j-1})\ldots \exp(-\imath \widehat{\theta_1} {B}_1)\Big |\Psi^{(0}\Big \rangle
\end{align*}}

\State {\textbf{(ii)}~ \textbf{Solve} the minimisation problem 
\begin{align*}
    \theta_j:= \argmin_{\theta \in [-\pi, \pi)} \mathcal{L}_j(\theta),
\end{align*}}

\State {\textbf{(iii)}~ \textbf{Update} $\widehat{\theta_j} \mapsto \theta_j$ and \textbf{update} $j\mapsto j+1$.}\\
\EndWhile
\end{algorithmic}
\end{algorithm}

Numerical testing of Algorithm \ref{alg:1} indicates that it may alleviate-- in some cases-- the issue of the GGA-VQE ansatz converging to a shallow local minimum. We refer, e.g., to Figure \ref{fig:GGA_2} above which displays results for a stretched, linear H$_6$ chain for which the GGA-VQE local optimisation ansatz performs poorly. However, for weakly correlated systems, the benefit of sequential reoptimisation in the presence of measurement noise is less clear. Given the additional measurement overhead needed to perform reoptimisation, we feel that this is best treated as a post-processing step to be performed only if resources allow.

\subsection{Additional Comparison Tests for GGA-VQE and ADAPT-VQE}

\begin{figure}[H]
    \centering
    \begin{subfigure}[b]{0.49\textwidth}
        \centering
        \includegraphics[width=\textwidth]{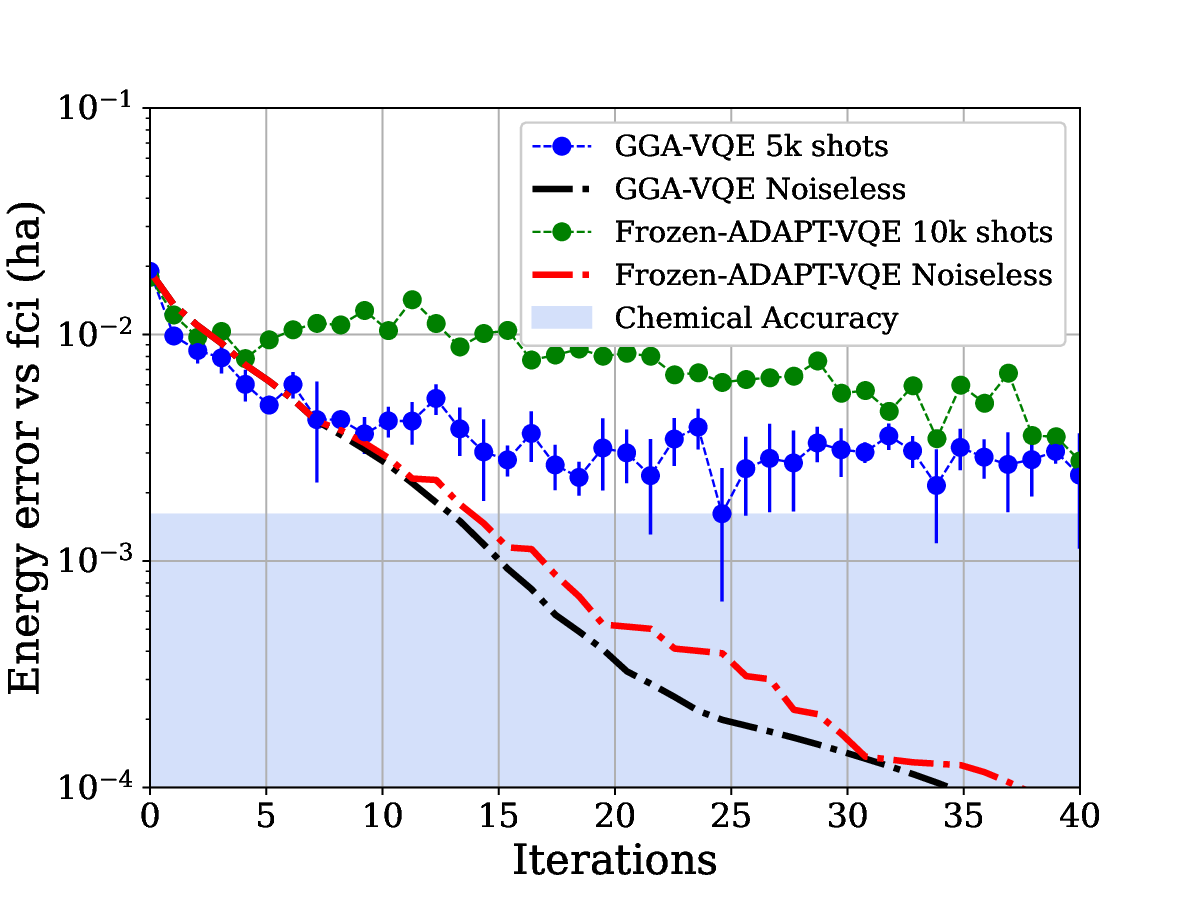}
        \caption{H\(_2\)O}
        \label{fig:h2oadapt}
    \end{subfigure}
    \begin{subfigure}[b]{0.49\textwidth}
        \centering
        \includegraphics[width=\textwidth]{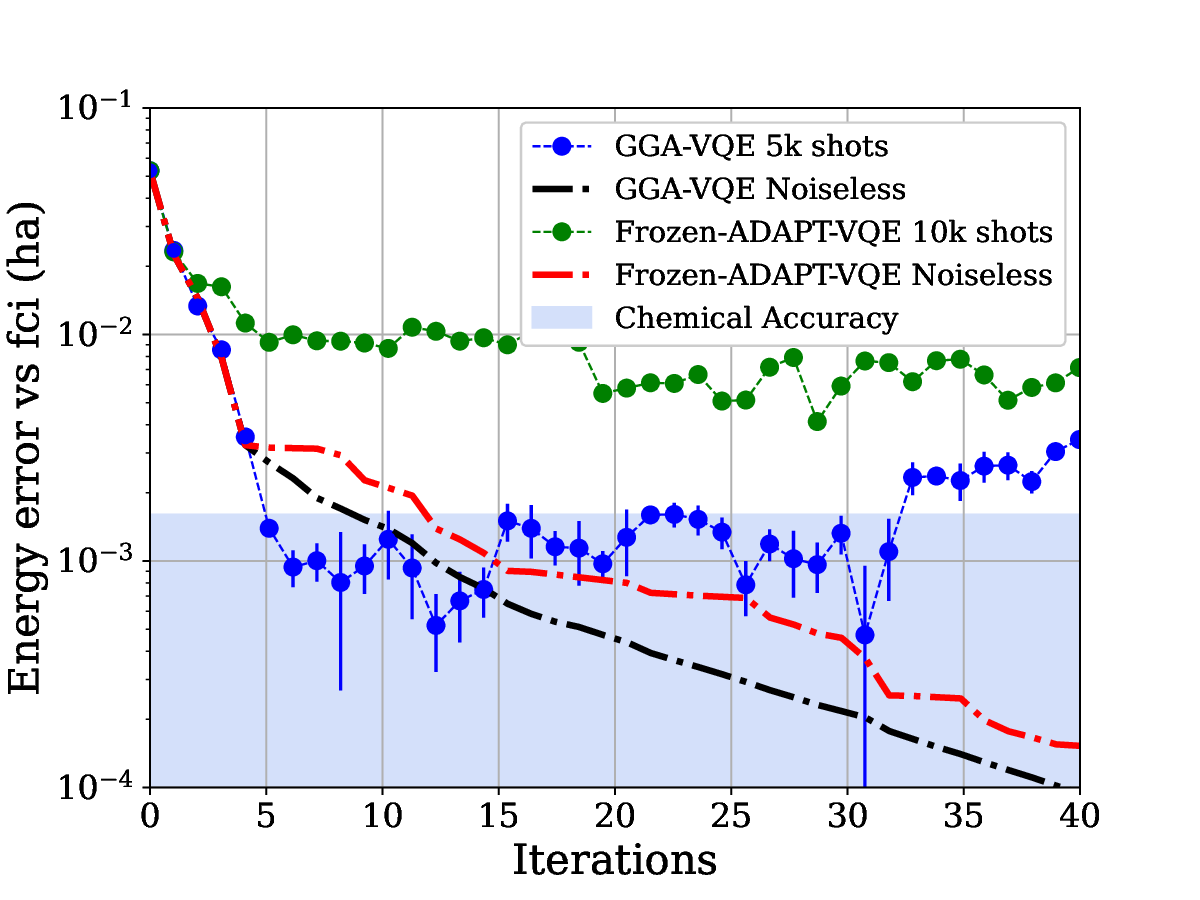}
        \caption{LiH}
        \label{fig:lihadapt}
    \end{subfigure}
    \caption{Comparison of the GGA-VQE algorithm with the Frozen-ADAPT-VQE algorithm, where only the last parameter is optimised and the core ansatz is kept frozen, for the ground state energy of H$_2$O and LiH. The plots represents the energy convergence as a function of the number of iterations of the algorithms. The shaded blue region indicates chemical accuracy at $10^{-3}$ Hartree. The numerical parameters used for these experiments were identical to those used for the tests in Section \ref{sec:numerics_1}.} 
    \label{fig:greedy}
\end{figure}

\begin{figure}[H]
    \centering
    \includegraphics[width=0.5\textwidth]{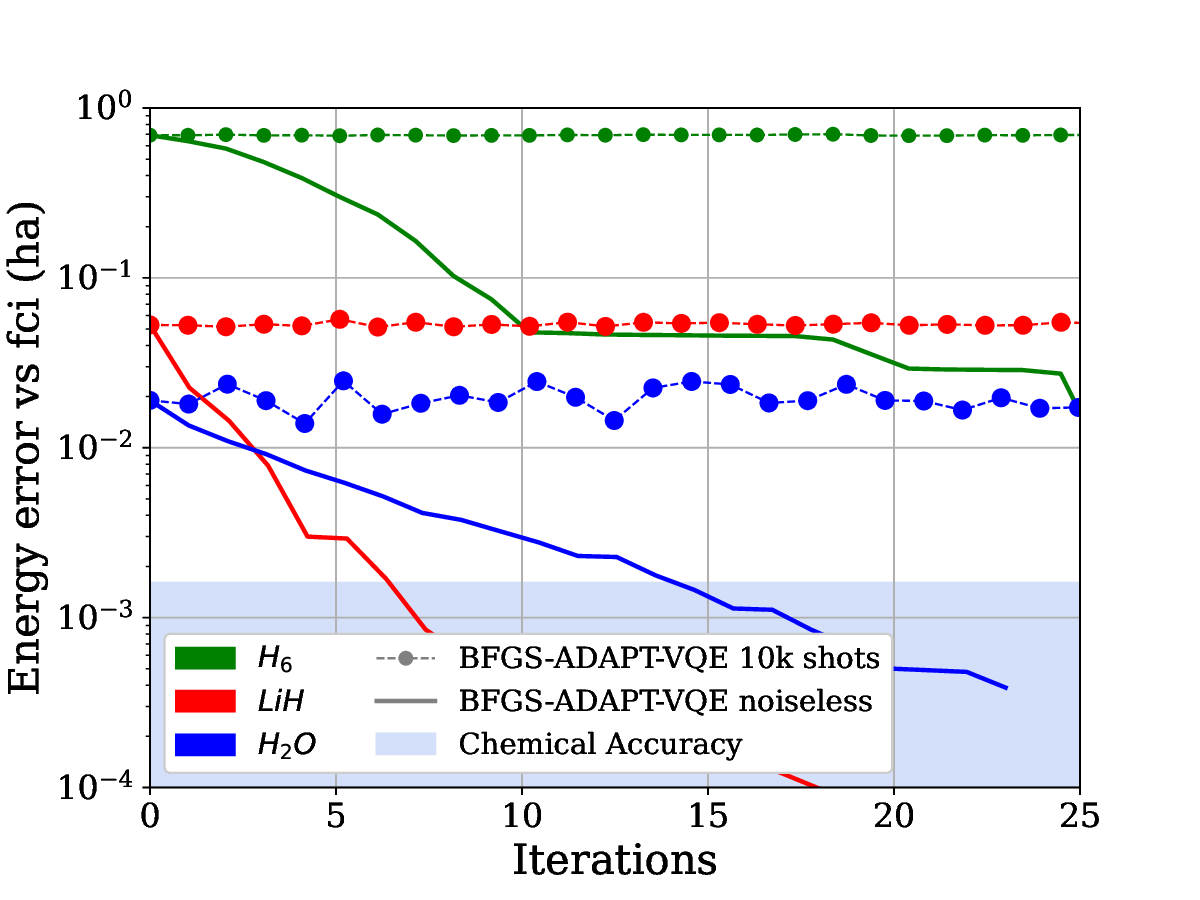}
    \caption{Comparison of ADAPT-VQE algorithm using the BFGS optimiser on the VQE subroutine under noiseless and shot noise conditions. The plots represents the energy convergence of LiH, H$_6$ and H$_2$O systems, as a function of the number of iterations of the algorithms. The shaded blue region indicates chemical accuracy at $10^{-3}$ Hartree.}
    \label{fig:bfgs}
\end{figure}

An additional question worth exploring is how the GGA operator selection and gradient-free analytic optimisation strategy compares with the gradient-based approaches used in ADAPT-VQE. To answer this question, we consider the weakly-correlated H$_2$O and LiH molecules previously studied in Section \ref{sec:numerics_1}, and apply a so-called "Frozen-ADAPT-VQE" algorithm to approximate their ground states. The "Frozen-ADAPT-VQE" method is a simplified version of the classical ADAPT-VQE in which the global optimisation step (see, e.g., Equation \eqref{eq:adapt_opt}) is replaced with a local optimisation step in which only the most recently appended parameterised unitary is optimised (using an iterative algorithm). The results of this numerical test-- in the same statistical noise framework considered in Section \ref{sec:numerics_1}-- are displayed in Figure \ref{fig:greedy}. The convergence results demonstrate that, even in the noiseless regime, the GGA operator selection procedure is more optimal than the ADAPT gradient-based selection criterion and produces a modestly improved ground state. On the other hand, in the presence of statistical noise due to observable measurement, we see that GGA-VQE outperforms ADAPT-VQE thus demonstrating the noise-resilience of our approach. Note that Frozen-ADAPT-VQE outperforms standard ADAPT-VQE simulations displayed in Figure \ref{fig:GGA_1}, in the explored noise regime for LiH, despite sharing the same operator selection criterion. This is likely due to the fact that global optimisation in the presence of measurement (statistical) noise guides the wavefunction to a plateau not reached by local optimisation of the last added operator.

To evaluate the suitability of different optimisation methods for noisy simulations, we carried out our three characteristic simulations (ground state energy of LiH, H$_6$ and H$_2$O) using BFGS as the VQE subroutine optimizer within ADAPT-VQE, both with and without shot noise. As shown in Figure 13, BFGS failed to achieve energy improvements under noisy conditions, making it unsuitable as an optimizer for this study and prompting the choice of COBYLA instead.

\bibliography{sn-bibliography.bib}








\end{document}